\documentstyle[sprocl,epsf,amstex]{article}

\def\beh{\begin{equation}}
\def\eeh{\end{equation}}
\def\bea{\begin{eqnarray}}
\def\eea{\end{eqnarray}}
\newcommand{\beq}{\begin{equation}}
\newcommand{\eeq}{\end{equation}}
\newcommand{\be}{\begin{eqnarray}}
\newcommand{\ee}{\end{eqnarray}}

\newcommand{\ga}{\lower.7ex\hbox{$
\;\stackrel{\textstyle>}{\sim}\;$}}
\newcommand{\la}{\lower.7ex\hbox{$
\;\stackrel{\textstyle<}{\sim}\;$}}
\renewcommand{\vec}[1]{{\bf #1}}

\begin{document}

\sloppy
\ 

\vspace{-1cm}

\title{ASPECTS OF CHIRAL SYMMETRY}

\author{A.V. SMILGA}

\address{Dept. of Physics, University of Nantes,\\
2 rue de la Houssini\`ere BP92208 \\ 
44322 Nantes CEDEX 3, France}

\maketitle


\abstracts{We give a pedagogical review of implications of
chiral symmetry in QCD. First, we briefly discuss classical textbook
subjects such as the
axial anomaly, spontaneous breaking of the flavor-nonsinglet 
chiral symmetry,
formation of light pseudo-Goldstone particles, and their 
effective interactions.
Then we proceed to other issues. We
explain in some detail 
a recent discovery how to circumvent the Nielsen--Ninomiya's theorem 
and implement chirally symmetric fermions on the lattice. We touch upon 
such classical issues as the Vafa-Witten's theorem and 't Hooft's anomaly 
matching conditions. We derive a set of 
exact theorems concerning the dynamics
of the theory in a finite Euclidean volume and the behavior of the Dirac
spectral density. Finally, we  discuss
an imaginary world with a nonzero 
value of the vacuum angle $\theta$. }
  
\vspace{.5cm}

\tableofcontents

\newpage

\section{Classic Sagas}

\subsection{Basics}

When a field theory enjoys a symmetry, it is always a fortunate circumstance
because the presence of the symmetry always simplifies analysis and,
more often than not,
 allows one
to obtain exact or semi-exact results. The richer is the symmetry, the stronger
are the results thus obtained. 
Sometimes, the presence of a rich enough symmetry allows one to {\it solve} the
theory (the best example is the exact results for the
spectrum of  the ${\cal N}=2$ supersymmetric Yang--Mills theory due to Seiberg 
and Witten\,\cite{SW}). 
 The symmetry of QCD is not so rich and the theory is not exactly solvable.
Still, the presence of light quarks such that the theory enjoys 
chiral symmetry in the chiral limit, when the quarks become massless,
 allows  one to extract a lot of consequences concerning  QCD dynamics,
with some of them having the status of {\it exact theorems}. In this 
mini-review we will concentrate on some
 less known and/or comparatively recent  results obtained   in this direction.
To make the discussion coherent, we will review some well-known
facts as well.

The Lagrangian of QCD reads
\be
\label{LQCD}
 {\cal L}_{QCD}  = - \frac 1{2g^2} {\rm Tr} \{{F}_{\mu\nu} {F}^{\mu\nu} \}
 \ +\  \sum_{f=1}^6 \bar\psi_f (i{\cal D}\!\!\!\!/ - m_f) \psi_f 
\ +\ \frac {\theta}{32\pi^2} 
 F^a_{\mu\nu}  \tilde F^{a\ \mu\nu}\  ,
  \ee
where  $ \tilde F_{\mu\nu} = \frac 12
   \epsilon_{\mu\nu\alpha\beta} F_{\alpha\beta}$, the physical value
of the parameter $\theta$ is very small $\theta \la 10^{-9}$,  and 
$m_f$ are the quark masses. The $u$, $d$, and $s$ quarks are relatively
light, with masses $m_u \approx 4 \ {\rm MeV},\ m_d \approx 7\  {\rm MeV}$, and
$m_s \approx 150\  {\rm MeV}$. $m_{u,d}$ are especially small. It makes
sense to consider the  chiral limit when $N_f = 2$ or $N_f = 3$ quarks
become massless. In this limit, the Lagrangian (\ref{LQCD}) is invariant under
the transformations
 \be 
\label{symvect}
\delta \psi_f = i \alpha_A  [t^A \psi]_f
  \ee
and
  \be 
\label{symax}
\delta \psi_f = i \beta_A \gamma^5 [t^A \psi]_f
  \ee
where $t^A$ ($A = 0, 1, \ldots, N_f^2 - 1$) 
are the generators of the flavor U$(N_f)$ group. The 
symmetry (\ref{symvect}) is the  ordinary isotopic symmetry.
It is still present  even if the quarks are endowed with 
a mass (of the same magnitude for all
flavors). The symmetry (\ref{symax}) holds only in the massless theory. The 
corresponding Noether currents are
   \be 
(j^\mu)^A \ = \ \bar \psi t^A \gamma^\mu \psi, \ \ \ \ 
(j^{\mu 5})^A \ = \ \bar \psi t^A \gamma^\mu \gamma^5 \psi\,.
 \label{curvecax}
  \ee
They are conserved upon applying the classical equations of motion.

\subsection{Singlet Axial Anomaly}

Let us discuss first the singlet axial symmetry (with $t^A = 1$). 
An important fact is that this symmetry  exists only in the
classical case. The full quantum path integral {\it is} not invariant under
the transformations
\be
\delta \psi = i\alpha\gamma^5 \psi,\ \ \ \ \ \delta \bar\psi = i\alpha
\bar\psi \gamma^5\ ,
\label{U1chir}
  \ee
where the flavor index is now omitted. 
This explicit symmetry breaking due to quantum effects can be
presented as an   operator identity involving an anomalous divergence,
  \be
\label{chiranom}
\partial_\mu j^{\mu 5} \ =\ - \frac 1{8\pi^2}
\epsilon^{\alpha\beta\mu\nu} {\rm Tr} \{F_{\alpha\beta} F_{\mu\nu}\}\ .
  \ee
There are  many ways to derive and  understand this relation. 
Historically, it was first derived by purely diagrammatic methods. 
The amplitude
 \be
\label{T3point}
\delta^{ab} T_{\mu\nu\lambda}(k,q) \ =\ 
\int d^4x d^4y e^{ikx} e^{iqy} 
\langle T \{j^a_\mu(x) j^b_\nu(0) j_\lambda^5(y) \}  \rangle_0\,,
  \ee
  ($j_\mu^{a,b} = \bar \psi t^{a,b} \gamma_\mu \psi$, where $a,b$  
are color indices)
 cannot be rendered
transverse with respect to  both vector and axial indices such that
$k_\mu T_{\mu\nu\lambda} = 0$ {\it and} $q_\lambda T_{\mu\nu\lambda}  = 0$  
due to 
ultraviolet divergences in the   {\it anomalous triangle} graph.
We will comment more on this  later   when we will be discussing  't Hooft's
self-consistency conditions.

  The anomaly relation (\ref{chiranom}) can also be derived as an operator 
identity. To this end, 
one should carefully define the operator of the axial current as
  \be
\label{axSchw}
j^{\mu 5} = \lim_{\epsilon \to 0} \bar\psi(x+\epsilon) \gamma^\mu \gamma^5
\exp \left\{ i \int_{x}^{x+\epsilon} A_\alpha(y) dy^\alpha
 \right\} \psi(x)
\ee
and perform the limit, averaging over the directions of $\epsilon_\mu$
 and making use of the operator 
equations of motion (see e.g. Ref. 2 for more details.).

Let us dwell here on the third and the most
elegant way to derive the anomaly relation (\ref{chiranom}) which
uses   path
integral methods. The anomaly appears due to the necessity
to regularize  the theory in the ultraviolet. The 
most  politically  correct
 approach would be to study a path integral regularized
by a lattice, and we will do so in Sec.~2. 
 Let us,  first, describe
a more standard and habitual {\it finite mode} regularization.

Consider an Euclidean path integral for the partition function in QCD
with one massless quark flavor. The fermionic part of the integral is
 $$
\int \prod_x {d\bar \psi(x) d\psi(x)} \exp\left\{i\int d^4x \bar \psi
 /\!\!\!\!{
\cal D}^E \psi  \right\}\ , 
$$
which formally coincides with the determinant of the Euclidean 
Dirac operator 
$-i /\!\!\!\!{ \cal D}^E = -i\gamma_\mu^E {\cal D}_\mu$, where
$\gamma_\mu^E$ are Euclidean $\gamma$ matrices. 
The latter are anti-Hermitean and satisfy 
$$\gamma_\mu^E \gamma_\nu^E + \gamma_\nu^E \gamma_\mu^E \ = 
\ -2\delta_{\mu\nu}\,.$$ From now on, we will stay 
in the Euclidean space and will suppress the superscript
``$E$'' for $/\!\!\!\!{ \cal D}$ and $\gamma_\mu$. 

Let us assume that the theory is somehow regularized in the infrared so 
that the spectrum of the operator $ /\!\!\!\!{ \cal D}$ is discrete. 
Note now that  the spectrum 
of the massless Euclidean Dirac operator enjoys the following 
symmetry: for any eigenfunction 
$u_k$ of
the operator $/\!\!\!\!{\cal D}$ with  
an
eigenvalue $\lambda_k$, the 
function $u_k' = \gamma^5 u_k$ is also an eigenfunction with
 the eigenvalue $-\lambda_k$ ($/\!\!\!\!{\cal D}$ and $\gamma^5$ 
anticommute). Therefore,
  \be
  \label{det0n0}
\det \| -i /\!\!\!\! {\cal D} + m \| \ =\ 
\prod_k (-i\lambda_k + m) \ =\ m^q \prod_k' (\lambda_k^2 + m^2)\ ,
  \ee
where $q$ is the number of possible 
 exact zero modes  and the product $\prod_k'$ 
runs over the nonvanishing 
eugenvalues. We see that the expression (\ref{det0n0}) is real, and the
Euclidean partition function is real,  too.
 
Let
us expand $\psi(x),\ \bar \psi(x)$ as 
  \be
\label{ferEbasis}
\left\{
\begin{array}{c}
 \psi(x) \ =\ \sum_k c_k u_k(x)  \\[0.2cm]
\bar  \psi(x) \ =\ \sum_k \bar c_k u_k^\dagger(x)
\end{array} \right. \ ,
  \ee
where $\{ u_k(x) \}$ is a 
 a set of eigenfunctions of $/\!\!\!\!{\cal D}$ forming a
complete basis in the corresponding Hilbert space.
   Then
  \be
\label{dcn}
 \prod_x {d\bar \psi(x) d\psi(x)} \ \equiv \ \prod_k d \bar c_k dc_k\,.
 \ee
Suppose the field variables are transformed by an infinitesimal
global chiral transformation (\ref{U1chir}).\footnote{We 
keep the Euclidean $\bar \psi$ and 
$\psi$ independent and are not bothered by the fact that 
$\delta \bar \psi \neq
(\delta \psi)^\dagger$.} Now,
 $\psi' = \psi + \delta \psi$ and  
$\bar \psi' = \bar \psi + \delta \bar \psi$
 can be again  expanded in the series (\ref{ferEbasis}). The
new expansion coefficients are related to the old ones,
  \be
\label{cnprim}
c_k' \ =\ c_k + i\alpha \sum_m c_m \int d^4x \ 
u_k^\dagger (x) \gamma^5  u_m(x)
\ \equiv \sum_m(\delta_{km} + i\alpha A_{km}) c_m \nonumber \\[0.1cm]
\bar c_k' \ =\ \bar c_k + i\alpha \sum_m \bar c_m \int d^4x \ 
u_m^\dagger (x) \gamma^5  u_k(x)
\ \equiv \sum_m  \bar c_m (\delta_{km} + i\alpha A_{mk})\,. 
\ee
The point is that the transformation (\ref{cnprim}) has a nonzero Jacobian. 
We have
 \be
\label{JacC}
 \prod_k {d\bar c_k}' dc_k' \ =\ J^{-2}  \prod_k d\bar c_k dc_k\ ,
 \ee
where
\be
\label{JviaTr}
J \ =\ {\rm det} (1 + i \alpha A) \ \approx \ 
\exp \left\{ i \alpha \sum_k A_{kk} \right\}\,.
 \ee
Or, in other words,
 \be
\label{lnJ}
\ln J \ =\ i\alpha \int d^4x \sum_k u^\dagger_k(x) \gamma^5 u_k(x) 
\ + o(\alpha)\,.
\ee
One's first (wrong!) impression might be that  $ \int d^4x 
\sum_k u^\dagger_k(x) \gamma^5 u_k(x) $ is just zero. Indeed, 
as was already mentioned above,
the function $u_k' = \gamma^5 u_k$ is also an eigenfunction of the
Dirac operator  with the
  eigenvalue $-\lambda_k$. { If} $\lambda_k \neq 0$, $u_k(x)$ and 
$\gamma^5
u_k(x)$ thereby represent  {\it different} eigenfunctions and the integral
$ \int d^4x \ u^\dagger_k(x) \gamma^5 u_k(x) $ vanishes.

A nonzero value of (\ref{lnJ}) is due to the fact that, for intricate
enough {\it topologically nontrivial} gauge fields, the spectrum of the 
Dirac operator involves some number of exact zero modes, for which $\gamma^5 
u_0(x) = \mp u_0(x)$ (depending on whether the modes are left-handed
or right-handed), and their contribution to the sum (\ref{lnJ}) is 
responsible for the whole effect. A famous theorem of Atiyah and Singer 
dictates
  \be
\label{AtSing}
 \int d^4x \sum_k u^\dagger_k(x) \gamma^5 u_k(x)  = n_R^{(0)} -
n_L^{(0)} \ =\ q \ ,
 \ee
where $n_{L,R}^{(0)}$ is the number of the left--handed (right--handed)
zero modes and $q$ is the topological charge  of the gauge
field configuration,
  \be
    \label{k4F}
     q \ =\ \frac 1{16\pi^2}  \int
   d^4x \ {\rm Tr} \{F_{\mu\nu} \tilde F_{\mu\nu}\}\,.
   \ee
  
Substituting (\ref{AtSing}) in Eqs. (\ref{lnJ}, \ref{JacC}), we now
see  that the change of the measure under the chiral transformation
can be presented as a shift of the effective action
  \be
\label{althet}
\delta S \  = \ \frac {i\alpha}{16\pi^2} \epsilon_{\alpha\beta\mu\nu} 
\int d^4x {\rm Tr} \{ F_{\alpha\beta} F_{\mu\nu}\,.
 \}
 \ee
In other words, a global singlet chiral transfomation 
is equivalent to leaving the fermionic fields intact, but  shifting instead
the parameter $\theta$ in the original theory  (\ref{LQCD}),
$$\theta \to \theta + 2\alpha\,.$$

\subsection{Spontaneous Breaking of Nonsinglet Chiral Symmetry}

Consider now the whole set of symmetries (\ref{symvect}), (\ref{symax}).
It is convenient  to introduce
 \be
 \label{psiLR}
\psi_{L,R}  = \ \frac 12 (1 \mp \gamma^5) \psi, \ \ \ \ 
\bar\psi_{L,R}  = \ \frac 12 \bar\psi (1 \pm \gamma^5)\ 
 \ee
and rewrite (\ref{symvect}), (\ref{symax}) as 
   \be
\label{chirtrans}
\psi_{L} \ \to \ V_L \psi_{L}, \ \ \ \ 
\psi_{R} \ \to \ V_R \psi_{R}\ , 
  \ee
where $V_L$ and $V_R$ are two different U$(N_f)$ matrices. The singlet
axial transformations with $V_L = V_R^* = e^{i\phi}$ are anomalous by the
same token as in the theory with a single quark flavor. Therefore, the true
fermionic symmetry group of  massless QCD is
  \be
\label{SULSUR}
{\cal G} \ = \ {\rm SU}_L(N_f) \times {\rm SU}_R(N_f) \times {\rm U}_V(1)\,.
 \ee
A fundamental {\it experimental} fact is that the symmetry (\ref{SULSUR})
is actually {\it spontaneously broken}, which means that the vacuum state
 is not invariant under the action of the group ${\cal G}$. The symmetry
${\cal G}$ is, however, not broken completely. The vacuum is still invariant 
under 
transformations  
with $V_L = V_R$, generated by the vector isotopic current.\footnote{An 
exact {\it theorem}\,\cite{Vafa}
that the vector symmetry cannot be
 broken spontaneously in QCD will be proven in Sec. 3.}
 
Thus, the pattern of breaking is
  \be
\label{patbreak}
 {\rm SU}_L(N_f) \times {\rm SU}_R(N_f) \ \to  {\rm SU}_V(N_f)\,.
  \ee
The nonvanishing  vacuum expectation values
 \be
\label{condfg}
\Sigma^{fg}\ = \ \langle  \psi_{L}^f  \bar \psi_{R}^g \rangle_0\ 
 \ee
are the order parameters of the
spontaneously broken axial symmetry. The matrix $\Sigma^{fg}$
is referred to as
the {\it quark condensate} matrix.

Non-breaking of the vector symmetry implies that the matrix order parameter
(\ref{condfg}) can be cast in the form
   \be
\label{conddel}
\Sigma^{fg}\ = \  \frac 12\, \Sigma \delta^{fg}
 \ee
by  group transformations (\ref{chirtrans}).
This means  that the general condensate matrix $\Sigma^{fg}$ is a 
unitary SU$(N_f)$ matrix multiplied by $\Sigma$. 

In general, $\Sigma$ could be any complex number. It can be
made real by a global U$_A(1)$ rotation which, according to Eq.~(\ref{althet})
(with the factor $N_f$),
amounts to a shift of the vacuum angle $\theta$. For {\it massless}
 quarks, physics
does not depend on the phase of $\Sigma$ and, thereby, on $\theta$.
 It is convenient then
to choose $\Sigma$ real and positive and $\theta = 0$. 
From  experiment, we know that  $\Sigma
 \approx (240 \ {\rm MeV})^3/2$ with about 30\% uncertainty (this value refers
to some particular normalization point $\mu \sim 0.5 \ {\rm GeV}$ on which  the
operator $\bar\psi \psi$ and its vacuum expectation value depend).

By  Goldstone's theorem, spontaneous  breaking
of a global continuous symmetry leads to the appearance of  
 purely massless Goldstone bosons. Their number coincides with the
number of broken generators, which is $N_f^2 - 1$ in our case.
 As it is the axial symmetry
which is broken, the Goldstone particles are pseudoscalars. They are nothing
but pions for $N_f = 2$ or the octet $(\pi, K, \bar K, \eta)$ for $N_f = 3$.

It is a fundamental and important fact  that 
spontaneous breaking of  continuous symmetries not only creates
 massless Goldstone particles, but also fixes the 
{\it interactions} of the latter at low energies.
To see this, let us first recall  that the Goldstone field describes 
fluctuations of the order parameter, $$\Sigma^{fg} \to \Sigma^{fg}(x) =
\frac 12 \Sigma U(x) $$ with $U(x) \in SU(N_f)$.
 It is convenient to express 
$U(x)$ as an exponential
  \be
 \label{Uchirexp}
U(x) \ =\ \exp \left\{  \frac {2i \phi^a(x) t^a}{F_\pi} \right \}\ ,
 \ee
where $\phi^a(x)$ are the physical meson fields [so that $\phi^a(x) = 0$ 
corresponds
to the vacuum (\ref{conddel})] and $F_\pi$ is a constant of dimension of mass.

The Goldstone particles are massless whereas all other states 
in the physical
spectrum have  nonzero mass. Therefore, we are  in the 
{\it Born-Oppenheimer} situation: there are two distinct energy scales and 
one can
 write down an effective Lagrangian depending only on {\it slow} Goldstone
 fields
with the {\it fast} degrees of freedom corresponding to all other particles 
being integrated out. 

To fix the exact form of this Lagrangian, note that the transformations
(\ref{chirtrans}) are realized at the level of the effective Lagrangian as
$U \to V_L U V_R^\dagger$. Any scalar function depending on $U$ and invariant
under this symmetry is also a function of $U^\dagger U = 1$, i.e. it is just
a constant. There is only one invariant structure involving two derivatives,
namely
 \be
\label{Leffchir2}
{\cal L}_{\rm eff}^{(2)} \ =\ \frac {F_\pi^2}4 {\rm Tr} \left\{ \partial_\mu U
 \partial^\mu U^\dagger \right \}\,.
 \ee
Take $N_f = 2$. The  perturbative expansion of Eq.(\ref{Leffchir2}) 
in powers of $\phi$ reads
   \be
\label{Lexpphi}
{\cal L}_{\rm eff}^{(2)} \ =\ \frac 12 ( \partial_\mu \phi^a )^2 +
\frac 1{6 F_\pi^2} \left[ (\phi^a  \partial_\mu \phi^a )^2 - (\phi^a \phi^a)
(\partial_\mu \phi^b)^2 \right] + \ldots
 \ee
Also  for $N_f \geq 3$, we have, on top of the standard kinetic term,
 a quartic
term involving two derivatives, with somewhat   more 
complicated group structure.
We see that the  symmetry dictates  rather specific interactions between
 the pions. They do not interact at the $s$-wave level, 
which means that all amplitudes vanish at zero momenta, 
but the strength of
interaction grows rapidly with energy. 

Equation (\ref{Leffchir2}) describes the 
effective chiral Lagrangian of massless Goldstone bosons at leading order. 
The first corrections 
involve 4 derivatives and there are 3 different invariant functions
   \be
\label{Leffchir4}
{\cal L}_{\rm eff}^{(4)} &=& L_1 
 {\rm Tr} \left\{ \partial_\mu U \partial^\mu U^\dagger \right \}^2 
+ L_2 {\rm Tr} \left\{ \partial_\mu U \partial_\nu U^\dagger \right \}
{\rm Tr} \left\{ \partial^\mu U \partial^\nu U^\dagger \right \}  
\nonumber \\[0.2cm]
&+& L_3  {\rm Tr} \left\{ \partial_\mu U \partial^\mu U^\dagger 
 \partial_\nu U \partial^\nu U^\dagger \right\} 
 \ee
(only 2 linearly independent structures are left for $N_f = 2$). 

The relevant Born-Oppenheimer expansion parameter is $\kappa_{\rm chir}
\sim p^{\rm char}/F_\pi$. When  $\kappa_{\rm chir} \sim 1$ (in practice, one
should rather take  $\kappa_{\rm chir} \sim 2\pi$), the  Born-Oppenheimer
approximation as well as  the whole effective Lagrangian approach
 breaks down, and
non-Goldstone degrees of freedom become important. The physical
meaning of $F_\pi$ is thus clarified. It characterizes the gap in the
spectrum and sets a scale below which massive degrees of freedom can be 
disregarded.

Let us discuss the actual world now. The  Lagrangian of real QCD 
(\ref{LQCD}) is 
not invariant under the axial symmetry transformations just because quarks
have nonzero masses. The symmetry (\ref{SULSUR}) is still very much relevant 
to QCD because {\it some} of the quarks happen to be very light. 

For $N_f = 2$, spontaneous breaking of an 
exact ${\rm SU}_L(2) \times {\rm SU}_R(2)$ symmetry would lead
to the existence of 3 strictly massless pions. As the symmetry is 
not quite exact, the pions have a small mass. However, 
their mass $M$ goes
to zero in the chiral limit $m_{u,d} \to 0$.
Indeed, trading the mass term
 \be
- m_u \bar u u - m_d \bar d d = m_u (u_L \bar u_R +  u_R \bar u_L)
+ m_d (d_L \bar d_R +  d_R \bar d_L)
 \ee
in the QCD Lagrangian for the contribution 
 \be
 \label{Lchirmass}
{\cal L}_{\rm eff}^{(m)} \ =\ 
\Sigma \ {\rm Re} \left[ {\rm Tr} \{ {\cal M} U^\dagger \} \right] 
  \ee
(${\cal M}$  is the quark mass matrix
which is chosen here in the form  ${\cal M} = {\rm diag} (m_u, m_d)$ with
real $m_u, m_d$)
in the effective chiral Lagrangian\,\footnote {Equation
 (\ref{Lchirmass}) is the leading chiral-noninvariant 
contribution
in ${\cal L}_{\rm eff}$ . Also  terms of higher order in ${\cal M}$, as well
as  terms of  first order in ${\cal M}$ but involving derivatives of $U$,
are allowed.}
 and
expanding (\ref{Lchirmass}) in $\phi^a$, we obtain the 
{\it Gell-Mann-Oakes-Renner} relation
  \be
 \label{GMOR}
F_\pi^2 M^2  =  (m_u + m_d)\Sigma + O(m_q^2)\,.
 \ee
  The  constant $F_\pi$ appears also in  the matrix
element $$\langle {\rm vac} | A_\mu^+ |\pi \rangle_p = i \sqrt{2} F_\pi 
p_\mu^\pi$$ of the
axial current $A_\mu^+ = \bar d \gamma_\mu \gamma^5 u$   and determines the 
charged pion decay rate.  Experimentally, $F_\pi \approx 93 $ MeV.

The effective chiral Lagrangian involving the lowest (\ref{Leffchir2}),
 (\ref{Lchirmass}), and higher order terms   
(in addition to the terms of higher order in derivatives as in 
Eq.~(\ref{Leffchir4}),
there are also terms of higher order in mass) can be and was used to 
calculate the amplitudes of different processes involving pseudoscalar mesons
at low energies. The corresponding technique is called 
{\it chiral perturbation theory}.\cite{CPT}

\section{ Quarks on Euclidean Lattice}

The material of the previous section is well known and can be found in many
textbooks. 
We are in a position now to discuss less known issues and will start with 
the question of how to put quarks on the lattice. This issue 
 looked very  confusing for a long time, and
 has  been clarified only recenly. 

\subsection{The Nielsen--Ninomiya's No-go Theorem}

The only way to {\it define} what  quantum field theory in nonperturbative
regime really 
means is to introduce a lattice ultraviolet regularization so that the symbol
of path 
integral is defined as a limit of finite-dimensional lattice integrals
when the lattice spacing goes to zero. We want the regularized theory to
preserve as many symmetries which the original continuous theory has as
possible. Actually, some symmetries can be broken on the lattice
(as rotational and Lorentz  symmetries are) and be restored in the
 continuum limit, but it is not always straightforward and not always
{\it true} that the symmetries are 
restored, indeed. For example, breaking gauge symmetry on the lattice would
probably lead to a nonsensical,  not gauge-invariant continuum theory.
Gauge-invariant lattice action for a pure gauge theory was written by Wilson
long time ago. But the task to put {\it fermions} on the lattice so that not
only the gauge symmetry, but also {\it chiral} symmetry would be 
preserved turned
out to be much more difficult.

Let us define to this end Grassmann
 variables $ \bar \psi_n, \ \psi_n$ in the nodes of the lattice for each
 quark flavor (color and Lorentz indices are not displayed). As a first 
and natural guess, let us write the lattice counterpart of the Dirac
 action as follows
  \be
  \label{fermlat}
  S^{\rm ferm. lat.}  &=&  -\frac{ia^3}2 \sum_{n, \mu}
  [ \bar\psi_n U_{n, n+e_\mu} \gamma_\mu \psi_{n + e_\mu}
  -  \bar \psi_n U_{n, n-e_\mu} \gamma_\mu \psi_{n - e_\mu}]\nonumber\\[0.1cm]
 &+& ma^4 \sum_n
  \bar \psi_n \psi_n \ .
 \ee

We see that the action (\ref{fermlat}) reproduces the Euclidean  action
   \be
\label{SferE} 
S^{\rm ferm} \ =\ - \int d^4x [i \bar \psi \gamma_\mu ( \partial_\mu - 
i  A_\mu ) \psi  - m  \bar \psi \psi]
   \ee
 in the continuum limit. Indeed, for  free fermions
 $$ - \frac{ia^3}2 \sum_{n, \mu}
   \bar \psi_n  \gamma_\mu [\psi_{n + e_\mu}
  - \psi_{n - e_\mu}] \ \to \ - i \int d^4x 
  \bar \psi \gamma_\mu  \partial_\mu  \psi \,. $$
  Expanding  $U_{n , n + e_\mu} \equiv 1 - i a  A_\mu e_\mu 
+ O(a^2)$, we also restore  the
interaction term, and the last term in Eq. (\ref{fermlat}) turns into the
continuum mass term.
  The action (\ref{fermlat} ) is invariant under  the
gauge
   transformations when the $U$ and $\psi$ are transformed according to 
\be
 \label{gaugelat}
 U_{n + e_\mu, n} \ &\to& \ \Omega_{n + e_\mu} U_{n + e_\mu, n} 
\Omega^\dagger_n
\ , \nonumber\\[0.2cm]
 \psi_n \ &\to& \ \Omega_n \psi_n\ ,
 \ee
 where $\{\Omega_n\}$ is a set of unitary matrices defined on the nodes of 
 the lattice. 

Equation ( \ref{fermlat}) is called the ``naive 
lattice fermion action'', and I have
to say that, if the reader was convinced by the above reasoning  that, in the
continuum limit, it goes over to Eq. (\ref{SferE}),  he/she was naive, 
too. Our implicit assumption was that the fermion fields $\psi_n$ depend on
the lattice node $n$ in a smooth manner, so that the finite difference
$\psi_{n + e_\mu} - \psi_{n - e_\mu}$ goes over to the continuum derivative.
It turns out, however, that  fermion field configurations which behave as
$\psi_n \sim (-1)^{n_1}$ or  $\psi_n \sim (-1)^{n_2 + n_4}$ and change 
significantly at the microscopic lattice  scale,  are equally
important. After carefully performing the continuum limit, these wildly 
oscillating modes give rise to 15 extra light fermion species with the same 
mass, the so-called {\it doublers}. 

To understand it, consider first free massless fermions. Let 
 \be
\label{derpernaz}
(\partial_\mu^+ \psi)_n \ =\ \frac 1a [ \psi_{n + e_\mu} - \psi_n ],\ \ \ 
(\partial_\mu^- \psi)_n \ =\ \frac 1a [ \psi_{n } - \psi_{n-e_\mu} ]
   \ee
be the forward and backward lattice derivative operators. The naive free 
massless Dirac operator is 
  \be
  \label{Dirlatnai}
{\mathfrak D}_{\rm free}^0 \ =\ - \frac i2 \gamma_\mu (\partial_\mu^+ 
+ \partial_\mu^-)\ .
  \ee
The eigenfunctions of ${\mathfrak D}_{\rm free}^0$ are characterized by the
Euclidean 4-momentum $p_\mu$,
$$ u_p(n) \ =\ C_p e^{ia  p_\mu n_\mu } \ ,$$
where $C_p$ is a constant Grassmann bispinor. The eigenvalue equation
${\mathfrak D}_{\rm free}^0  u_p(n) = -i \lambda_p  u_p(n) $ implies
 \be
\label{momlateig}
\left[ \frac 1a \gamma_\mu \sin (a p_\mu) \right] C_p \ =\   -i \lambda_p C_p
 \ee
with 
 \be
 \label{latfrspec}
\lambda_p \ =\ \pm \frac 1a \sqrt{\sum_\mu \sin^2 (a p_\mu)}\,.
  \ee
(The operator (\ref{Dirlatnai}) is anti-Hermitean and its eigenvalues are
purely imaginary.) 

When $ap_\mu \ll 1$, we reproduce the continuum massless fermions with the
spectrum $\lambda_p = \pm \sqrt{p_\mu^2}$. Each eigenvalue 
(\ref{latfrspec}) is doubly degenerate due to 2 possible polarizations. The 
eigenfunctionss with negative $\lambda_p$ are obtained from the ones with
 positive
$\lambda_p$ by multiplication by $\gamma^5$.

Note, however, that the lattice Dirac equation (\ref{momlateig}) has an
{\it additional} discrete symmetry $(Z_2)^4$: for any eigenfunction 
$u_p$, the function $\hat Q_\mu u_p 
\equiv \gamma_\mu \gamma^5 u_{p + (\pi/a)e_\mu}$
(no summation over $\mu$ ) is also the eigenfunction of 
${\mathfrak D}_{\rm free}^0$ with the same eigenvalue $\lambda_p$. The 
operators $\hat Q_\mu$ commute with ${\mathfrak D}_{\rm free}^0$ and 
anticommute
with each other 
$$\hat Q_\mu \hat Q_\nu + \hat Q_\nu \hat Q_\mu = 
2 \delta_{\mu\nu}\,.$$ 
The functions
 \be
\label{16plet}
u_p,\ \ \hat Q_\mu u_p, \ \ \hat Q_{[\mu} \hat Q_{ \nu ]} u_p, \ \ 
\hat Q_{[\mu} \hat Q_\nu \hat Q_{ \lambda ]} u_p,\ \ 
\hat Q_{[\mu} \hat Q_\nu  \hat Q_{ \lambda} \hat Q_{ \rho ]} u_p
 \ee
form a degenerate 16-plet. 

In the free case, each eigenstate of the naive Dirac operator is not just
16-fold, but 32-fold degenerate due to polarizations. In the interacting case
(on a generic gauge field background), polarization is not 
a good quantum number, but the 16-fold degeneracy (\ref{16plet}) still
 holds. The naive
lattice Dirac operator  in Eq.~(\ref{fermlat}) which
 can be written in the form 
${\mathfrak D}^0 = - \frac i2 \gamma_\mu ({\cal D}_\mu^+ +
{\cal D}_\mu^- ) $, where
\be
\label{dercovlat}
({\cal D}_\mu^+ \psi)_n \ =\ \frac 1a \left(\psi_{n + e_\mu}
U_{n, n+e_\mu} - \psi_n \right)\ , \nonumber \\ 
({\cal D}_\mu^- \psi)_n \ =\ \frac 1a \left(\psi_n  - 
\psi_{n - e_\mu} U_{n, n-e_\mu}  \right) \ 
  \ee
are the covariant lattice forward and backward derivatives,
still enjoys the symmetries
 \be
 \label{latdissym}
\hat Q_\mu : \ \psi_n \ 
\longrightarrow (-1)^{n_\mu} \gamma_\mu \gamma^5 \psi_n\ 
 \ee
(no summation over $\mu$).
We see that if $\psi_n$ changes smoothly from node to node, its 15 doublers
(\ref{16plet}) wildly oscillate on the microscopic lattice spacing scale.
We might call these modes ``unphysical'' but they would not listen to us and 
contaminate with a vengeance any numerical lattice calculation we might wish
to do. Some way to get rid of them should be suggested, otherwise QCD, the
theory involving only 6 quarks with different masses, would not be 
operationally defined.

The problem is that it is not so simple. Let us look for some other lattice 
Dirac operator ${\mathfrak D} \neq {\mathfrak D}^0$ satisfying the following
4 natural conditions:
 \begin{enumerate}
\item At distances much larger that the lattice spacing $a$, 
${\mathfrak D} \to -i/\!\!\!\!{\cal D}$ giving rise to a massless fermion in
the continuum limit.\footnote{Adding a 
finite mass term to ${\mathfrak D}$ presents no difficulties
[see Eq.~(\ref{fermlat})]. }

\item All the modes of ${\mathfrak D}$ not associated with the latter are of 
order $1/a$ (no doublers!). 

\item ${\mathfrak D}$ is local. In other words, the matrix elements
${\mathfrak D}_{n n'} $ decay exponentially fast at large distances
$|n - n'| \gg 1$. 

\item Chiral symmetries (\ref{U1chir}), (\ref{symax}) of the massless 
fermionic action are 
not broken
by the regularization  explicitly [the singlet axial symmetry  (\ref{U1chir})
 is eventually going to be
broken due to noninvariance of the  fermionic  measure, but we require
the absence of explicit breaking in the regularized Lagrangian]. This 
seems to imply the condition ${\mathfrak D}\gamma^5 +
\gamma^5 {\mathfrak D} = 0$.
 
\end{enumerate}

\begin{figure}
 \begin{center}
        \epsfxsize=200pt
        \epsfysize=200pt
        \vspace{-5mm}
        \parbox{\epsfxsize}{\epsffile{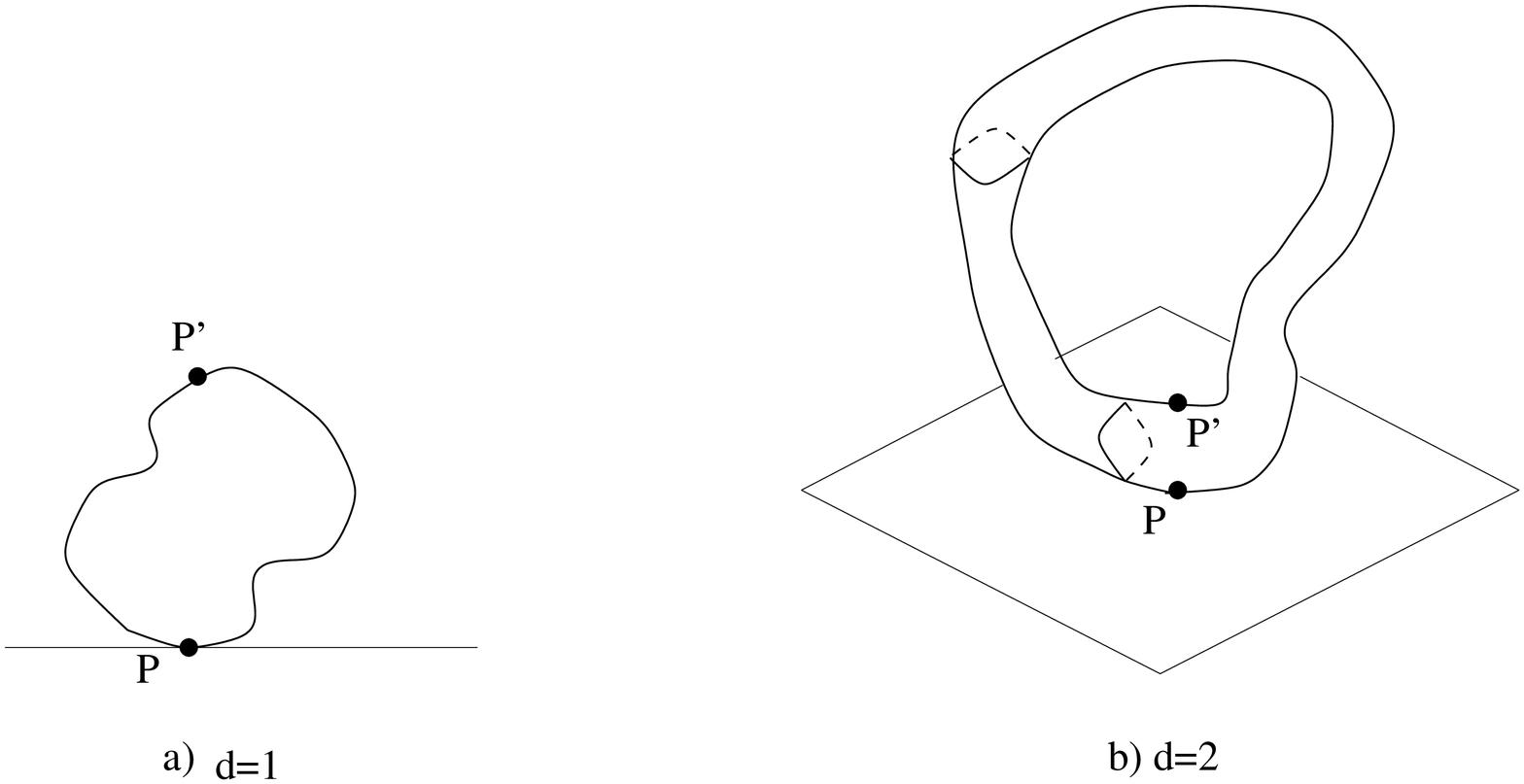}}
        \vspace{5mm}
    \end{center}
\caption{Nielsen--Ninomiya theorem. $P$ and $P'$ are different zeros of the
lattice Dirac operator.}
\label{proobr}
\end{figure}

The no-go theorem due to Nielsen and Ninomiya tells us, however, that such
${\mathfrak D}$ {\it does not exist}. To understand it, consider first 
the free fermion case. The momentum $p_\mu$ is then a good quantum number, and
the Dirac operator in the momentum representation has the form
 \be
{\mathfrak D} (p) \ =\ \gamma_\mu F_\mu (p) + G(p)\ .
 \ee
The condition 4 tells us that $G(p) = 0$. The condition 1 implies
that $F_\mu(p) = p_\mu + O(ap^2) $ for $ap_\mu \ll 1$ . Now,
$F_\mu (p)$ is a periodic function of its four arguments $p_\mu$ with the 
period $2\pi/a$. It realizes thus a smooth map $T^4 \to R^4$, where
$R^4$ is the tangent space. A look at Fig.~\ref{proobr}
can convince the reader that the point  on the tangent space where it touches
our torus has at least one more pre-image. And his/her intuition would not
betray him/her: 
this statement can be proven in a rigorous mathematical manner. 
Basically, it follows from the fact that the degree of the map 
 $T^d \to R^d$ is zero which means that\,\cite{DNF}
 \be
\label{degrmap}
\sum_{
\begin{array}{c}
{\small \rm  pre-images} \\{\small \rm  of}\ P \end{array}} {\rm sign}\ 
\left[ \det \|
\partial_\nu F_\mu(p) \| \right] \ =\ 0\ .
  \ee
 As the Jacobian of the mapping $p_\mu \to F_\mu(p)$ 
is equal to 1 at the point $p_\mu = 0$,
Eq.(\ref{degrmap}) implies that some other pre-images of zero, i.e. some other
solutions of the equation system $F_\mu (p) = 0$ should be present (one can
have just one extra solution as in Fig.~\ref{proobr} or more: for the ``round
upright torus'' $F_\mu(p) = \sin(ap_\mu)/a$, there are $2^d - 1$ extra
solutions). And that means
the presence of doublers in contradiction with condition 2.

The only remaining possibility is that $F_\mu (p)$ are not continuous. Besides
being ugly, it contradicts also  condition 3: the matrix elements
${\mathfrak D}_{n n'} = {\mathfrak D}(n - n') = \gamma_\mu F_\mu (n - n')$
present actually the Fourier coefficients of the periodic function
$\gamma_\mu F_\mu (p)$. If the latter is discontinuous, the Fourier
coefficients cannot decay faster than $1/|n - n'|$ (otherwise, the Fourier
series would converge uniformly on the torus $p_\mu \in \left[0, \frac{2\pi}a 
\right)$ and the 
sum of such a series would be continuous). 

We have proven that a lattice Dirac operator satisfying the above
four conditions
cannot be found for free fermions, but that also means that it cannot be 
found in QCD: any ${\mathfrak D}$ with this property should also enjoy it
 for any smooth set of the link variables\,\footnote{We 
{\it believe} that, in the continuum limit $a \to 0$ , the
characteristic fields $\{U\}$ contributing to the path integral can be 
{\it gauge  transformed} into the form where
$U = 1 + O(a)$ for all links. Note that the statement that
characteristic fields  {\it are}  always smooth  is just wrong: there are
instantons which, in the singular
gauge, involve singularities of the gauge potential 
$\sim x_\mu/x^2$ at the instanton center. Note also that 
a gauge transformation removing the singularity at $x = 0$ moves it to some 
other point or to infinity. For the Euclidean torus that means that 
we cannot {\it simultaneously} require $U_{\rm all\ links} = 1 + O(a)$
and periodicity of $U$.

   To the best of our knowledge, a 
rigorous proof of 
this  crucial assumption is absent, but it can be justified by arguing
that the action of field configurations which are ``essentially singular''
 (so that the singularity cannot be removed by a gauge transformation) would 
be infinite  in the continuum limit. } 
and, in particular, for the set $U_{\rm each\ link} = 1$ corresponding to
 free theory.

\subsection{Ways to Go. The Ginsparg-Wilson Way.}

If we still want to build up a lattice version of QCD, we have to relax
at least one of our four conditions. Conditions 1 and 2 are, however, 
indispensible: a lattice theory where they do not hold just has nothing to
 do with QCD. Therefore, either  locality  or chiral invariance of the lattice
action should be abandoned.

One of the possible procedures is that only {\it one mode} of each degenerate
16-plet of ${\mathfrak D}^0$ is taken into account in the 
fermionic determinant
and in the spectral decomposition of fermion Green's functions
\be
 \label{speclat}
\langle \psi_n \bar \psi_{n'} \rangle \ =\ \sum_{k}
\frac {u_k (n) u_k^\dagger (n') }{m - i \lambda_k}\ ,
 \ee 
etc,  where $u_k (n)$ describe the $k$-th eigenmode of  ${\mathfrak D}^0$ as a 
function of the node.  This amounts to choosing the lattice Dirac operator in 
the form
$({\mathfrak D}^0)^{1/16}$,  which is not local. A similar method is sometimes
used in the practical lattice calculations, but besides purely technical
inconveniences  it is 
unsatisfactory from a philosophical viewpoint: we {\it would} 
like to have a local lattice approximation for a local field theory.  

But then the chiral invariance (\ref{U1chir}) is necessarily lost. 
Though renouncing 
chiral invariance is also not desirable -- when regularizing the theory, we
should try to preserve as much of its symmetries as possible --  it is still
considered as the least of evils.

Two ways of chiral noninvariant lattice regularization have been known for
some time and used in practical calculations: {\it (i) Wilson} fermions and
 {\it (ii) Kogut-Susskind} or {\it staggered} fermions. We will describe
here the first method which consists in adding to ${\mathfrak D}^0$ the term
$\sim a {\cal D}_\mu^-  {\cal D}_\mu^+ =  a {\cal D}_\mu^+  {\cal D}_\mu^-$,
where the covariant lattice derivatives $  {\cal D}_\mu^\pm$ are defined in
Eq.~(\ref{dercovlat}). Thus, the Wilson-Dirac operator is defined as 
 \be
 \label{WilsDir}
{\mathfrak D}^W \ =\ - \frac i2 \gamma_\mu \left( {\cal D}_\mu^+ + 
{\cal D}_\mu^- \right) - \frac {ra}2 {\cal D}_\mu^+ {\cal D}_\mu^-
 \ee
where $r$ is an arbitrary nonzero real constant. For free fermions,
$ {\cal D}_\mu^\pm \to  \partial_\mu^\pm$ and
$$ ( \partial_\mu^+  \partial_\mu^- \psi)_n = \frac 1{a^2} \sum_\mu
\left( \psi_{n + e_\mu} +  \psi_{n - e_\mu} - 2 \psi_n \right) $$
which is the lattice laplacian. Passing to the momentum representation,
we obtain
 \be
\label{DWmom}
{\mathfrak D}^W_{\rm free} (p) \ =\ \frac 1a \gamma_\mu \sin (ap_\mu)
+ \frac {2r}a \sum_\mu \sin^2 \left( \frac {ap_\mu}2 \right)\,.
 \ee
The second term has the form of momentum-dependent mass. For small $p_\mu \ll
1/a$, it can be neglected and the continuum massless Dirac operator is
reproduced. In contrast to ${\mathfrak D}^0$, the operator (\ref{DWmom}) 
is not anti-Hermitean, and its eigenvalues are complex. What is important is
that, for not small $p_\mu$, the absolute values of the eigenvalues of 
 ${\mathfrak D}^W\,$,
    \be
    \label{eigWil}
 -i \lambda_p^W \ =\ \pm \frac ia  \sqrt{\sum_\mu \sin^2 (a p_\mu)}
  + \frac {2r}a \sum_\mu \sin^2 \left( \frac {ap_\mu}2 \right) 
   \ee
are of order $1/a$. The doublers disappear. At 
$p_\mu = \left( \frac \pi a, 0,0,0 \right)$, the eigenvalue 
(\ref{eigWil}) is $\frac {2r}a$; at 
$p_\mu = \left( 0, \frac \pi a, 0, \frac \pi a
 \right)$, it is $\frac {4r}a$, etc. 

The chiral symmetry is broken, however, and it is messy. In principle, when
the continuum limit $a \to 0$ is taken, the effects due to the breaking
of $\gamma^5$ invariance must be suppressed, but in this particular 
problem the continuum limit with restoration of chiral symmetry is rather
slow to reach, and it is even not shown {\it quite} conclusively that it is
reached at all. In particular, it is difficult to render pions light. In 
practical calculations, it is achieved by introducing 
a large bare quark mass of order $g^2(a)/a$ and fine-tuning it so that the 
effects due to
two chiral noninvariant terms -- the Wilson term and the bare quark term --
would cancel each other. Needless to say, this is a rather artificial
and unaesthetic procedure.

As we see, this Nielsen-Ninomiya puzzle defies attempts to solve 
it.
A recent remarkable observation is that the best strategy here is 
to follow the example of the Alexandre the Great and just cut it through!
The adequate sword was forged back in 1982 by Ginsparg and Wilson\,\cite{GW}. 
They
suggested to consider the lattice Dirac operators satisfying  the relation
 \be
 \label{GWrel}
\gamma^5 {\mathfrak D} + {\mathfrak D} \gamma^5 \ =\ a {\mathfrak D} \gamma^5
{\mathfrak D}\,.
  \ee
The anticommutator $\{ {\mathfrak D}, \gamma^5 \}$ does not vanish which means
that the lattice Lagrangian is not invariant with respect to chiral 
transformation
 \be
 \label{chirnai}
\delta \psi_n = i\alpha \gamma^5 \psi_n,\ \ \  \ \delta \bar \psi_n = 
i\alpha \bar \psi_n  \gamma^5 \ ,
 \ee
a lattice Euclidean counterpart of Eq.~(\ref{U1chir}).

It took 16 years to realize\,\cite{Lusch} that the lattice fermion action
 \be
 \label{SFgen}
S_F \ =\ a^4 \sum_{n n'} \bar \psi_n {\mathfrak D}_{n n'} \psi_{n'}
\ee
(color and spinor indices being suppressed), with ${\mathfrak D}$ 
satisfying the relation ({\ref{GWrel}), is invariant with respect to the
following
transformations:
 \be
 \label{transLu}
\delta \psi &=& i \alpha \gamma^5 \left[1 - 
\frac 12 a{\mathfrak D} \right] \psi
 \nonumber \\[0.1cm]
\delta \bar \psi &=& i \alpha \bar \psi  
\left[1 - \frac 12 a{\mathfrak D} \right] \gamma^5 \ .
 \ee 
If ${\mathfrak D}$ is local (in the sense of condition 3 in the 
Nielsen-Ninomiya list), Eq.~(\ref{transLu}) is as good a lattice approximation
of the continuous chiral symmetry (\ref{U1chir}) as the trivial 
(\ref{chirnai}).
In particular, the pions would automatically be light (massless in the chiral
limit), and no fine tuning is required. But  condition 4 above is no more
satisfied and one can hope now to find a local ${\mathfrak D}$
not involving doublers. The problem is still not trivial: as we will see a bit
later, many solutions of the Ginsparg-Wilson relation (\ref{GWrel}) can be
found and most of them  {\it are}
 not what we are looking for. The simplest {\it good} solution was suggested
by Neuberger.\cite{Neub} It has the form
 \be
\label{Neuberger}
{\mathfrak D} \ =\ \frac 1a \left[ 1 - \frac A {\sqrt{A^\dagger A}} \right]
 \ee
where $A = 1 - a {\mathfrak D}^W $ and ${\mathfrak D}^W$ is the Wilson--Dirac
operator (\ref{WilsDir}) with $r > 1/2$. In particular, for $r =1$ and for the
free fermions, we have 
 \be
\label{Neumom}
 a {\mathfrak D}(p) \ =\ 1 - \ \frac {1 - 2 \sum_\mu \sin^2 \left( \frac
{ap_\mu}2 \right) -  \gamma_\mu \sin (ap_\mu)}
{\left[ 1 + 8 \sum_{\mu < \nu} \sin^2 \left( \frac {ap_\mu}2 \right)
 \sin^2 \left( \frac {ap_\nu}2 \right) \right]^{1/2}}\ .
 \ee
 The eigenvalues of (\ref{Neumom}) are different from zero provided 
$p_\mu \neq 0$ . In particular, for $$p_\mu =  \left( \frac \pi a, 0,0,0 
\right)\,,\quad p_\mu =  \left( \frac \pi a, \frac \pi a ,0,0 \right)\,,\quad
p_\mu =  \left( \frac \pi a, \frac \pi a, \frac \pi a,0 \right)\,,$$
 and
$$
p_\mu =  \left( \frac \pi a, \frac \pi a, \frac \pi a ,  \frac \pi a 
\right)$$
the eigenvalues $-i\lambda$ of ${\mathfrak D}$ are all equal to $ 2/a$.
The doublers are absent. A second look at Eq. (\ref{Neumom}) reveals a 
beautiful feature displayed in Fig.~\ref{krugfig}: the eigenvalues of
${\mathfrak D}$ lie on the circle
 \be
 \label{krug}
\left( {\rm Re} \ \lambda \right)^2 +
\left( {\rm Im }\ \lambda - \frac 1a \right)^2 \ =\ \frac 1{a^2}\,.
 \ee
This property holds also in the interacting case. Using the property
${\mathfrak D}^\dagger = \gamma^5 {\mathfrak D} \gamma^5$ and the
Ginsparg-Wilson relation (\ref{GWrel}), it is not difficult to see that
the operator $V \ =\ 1 - a{\mathfrak D}$ is unitary, i.e. its eigenvalues lie
on the circle $\{ e^{i\phi} \}$. The eigenvalues of ${\mathfrak D}
= (1-V)/a$ lie on the circle in Fig.~\ref{krugfig}.

\begin{figure}
 \begin{center}
        \epsfxsize=200pt
        \epsfysize=200pt
        \vspace{-5mm}
        \parbox{\epsfxsize}{\epsffile{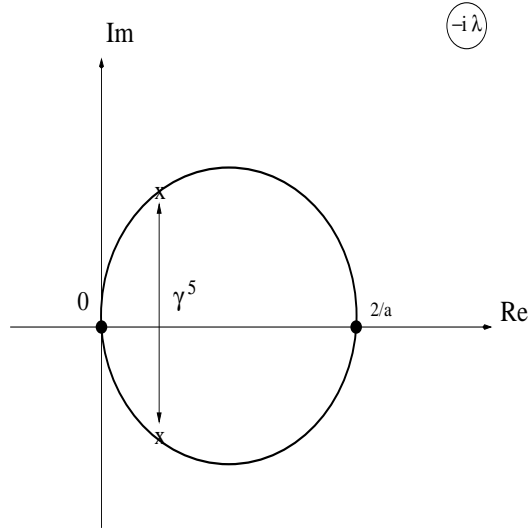}}
        \vspace{5mm}
    \end{center}
\caption{The circle of eigenvalues for  Neuberger's operator. The
eigenmodes with
the eigenvalues marked by crosses are related by a 
$\gamma^5$ transformation.}
\label{krugfig}
\end{figure}

The function (\ref{Neumom}) is analytic on the torus $p_\mu \in \left[ 0, 
\frac {2\pi}a \right)$ 
which means that its Fourier image decays exponentially at large distances.
The Dirac operator thus constructed is local. In the interacting case, it
stays local if the gauge field is smooth enough, i.e. the link variables 
$U_{n, n+e_\mu}$ are sufficiently close to 1. 

As was mentioned, the singlet axial symmetry (\ref{U1chir}) is anomalous which
shows up in the noninvariance of the fermionic measure. The measure
$\prod_n d\bar \psi_n d \psi_n$ is obviously invariant, however, with respect
to the ultralocal transformations (\ref{chirnai}). This follows from the
fact that ${\rm Tr}\{ \gamma^5\} = 0$. On the other hand, Eq.~(\ref{lnJ}) 
relates
the modification of the measure to the operator trace of $\gamma^5$,
 \be
\label{optrace}
 {\rm \bf Tr}\{\gamma^5 \} \ =\ \int d^4x \sum_k u^\dagger_k(x) \gamma^5 
u_k(x)
  \ee
 which is thus zero too. For the naive Dirac operator, only the
zero
modes contribute to the sum (\ref{optrace}) 
which means that the number of the right-handed and
left-handed zero modes of ${\mathfrak D}^0$ should be equal. That also
follows from the fact that ${\mathfrak D}^0$ commutes with $\hat Q_\mu$
defined in Eq.~(\ref{latdissym}). It is not difficult to see that if $u_k$
is, say, right-handed, 
$ \hat Q_{[\mu} \hat Q_{ \nu ]} u_k$  and
$\hat Q_{[\mu} \hat Q_\nu  \hat Q_{ \lambda} \hat Q_{ \rho ]} u_k$ are 
right-handed too, but $\hat Q_\mu u_k$ and 
$\hat Q_{[\mu} \hat Q_\nu \hat Q_{ \lambda ]} u_k$ are left-handed.
 Thus, instead of a single right-handed zero mode in 
(the lattice approximation
for) the instanton background, we have a degenerate 16-plet with 
8 right-handed and 8 left-handed modes,
which are no more necessarily zero modes. 
The vanishing of 
the index of  ${\mathfrak D}^0$ is closely related to 
the identity (\ref{degrmap}).
Indeed, the sign of the Jacobian $\det \|
\partial_\nu F_\mu(p) \|$ describes the orientation of a neighborhood of
the pre-image $P_i \in T^4$ with respect to the orientation of the tangent
space $R^4$ onto which it is mapped. This
orientation is obviously related to chirality.

The absence of the
anomaly is one of the diseases of the naive lattice Dirac 
operator. For the  operators of the  Ginsparg-Wilson type, the situation is 
different.
Chiral symmetry is now implemented as in Eq.~(\ref{transLu}) and,
generically, the measure {\it is} not invariant with respect to these 
transformations. We have, instead of Eq.~(\ref{lnJ})
 \be
\label{lnJGW}
\ln J \ =\ i\alpha {\bf Tr} \left\{ \gamma^5 \left( 1 - \frac 12 a
{\mathfrak D} \right) \right\}\,.
 \ee
Even though  {\bf Tr}$ \{ \gamma^5 \} = 0$ (we derived this in the basis 
involving the eigenvalues of ${\mathfrak D}^0$, but it is true in any 
basis), 
  {\bf Tr} $\{ \gamma^5 a {\mathfrak D}\}$  need not vanish and the anomaly 
is there. 

It is instructive to see how a nonzero operator trace in the
right-hand side of Eq.~(\ref{lnJGW}) is obtained in the basis 
involving the eigenvalues of ${\mathfrak D}$.
First, it is still true that for any eigenfunction $u_k$ of ${\mathfrak D}$,
$\gamma^5 u_k$ is also an eigenfunction with the eigenvalue
$\lambda_k' \ =\ -\lambda_k^*$ (complex conjugation appears and this is 
what distinguishes the Ginsparg-Wilson case from the continuum or
naive lattice Dirac operators). Thus, for almost all eigenstates on the circle
in Fig.~\ref{krugfig}), $u_k$ and $\gamma^5 u_k$ have {\it different}
eigenvalues, are orthogonal to each other,
 and the corresponding contribution  
vanishes. The only exception are two points on the circle:
$\lambda = 0$ and $\lambda = 2i/a$ where $\lambda_k = \lambda_k'$, and the
eigenstates can have a definite chirality. But the doublers 
$-i \lambda = \frac 2a$  obviously give
zero contribution in Eq.~(\ref{lnJGW}).
Only the zero modes of ${\mathfrak D}$ are relevant. We have derived the
{\it lattice index theorem}\,\cite{HLN}
 \be
 \label{indexlat}
n^0_R({\mathfrak D}) - n^0_L({\mathfrak D}) \ = \ 
- \frac 12 {\rm Tr} \{\gamma^5 a {\mathfrak D} \}\,.
 \ee 
The right-hand side
 of Eq.~(\ref{indexlat}) presents a functional depending on the link
variables $\{U\}$. By definition, it presents a sum over all lattice nodes
of some local expression. After some work, one can be convinced that it goes
over to the topological charge (\ref{k4F}) in the continuum limit.

\section{Continuum Theory: Some Exact Results}

We abandon the lattice now and will  discuss in this section 
various aspects of chiral symmetry in the continuum limit. Sometimes, we
will think in terms of quarks, of 
 symmetries (\ref{symax}) and  (\ref{U1chir}),
and of the order parameter (\ref{condfg}) associated with the spontaneous 
breaking of the flavor-nonsinglet symmetry. Sometimes, we will describe the
system in terms of the pseudo-Goldstone degrees of freedom and the effective
Lagrangians (\ref{Leffchir2}), (\ref{Leffchir4}), and  (\ref{Lchirmass}). 
Sometimes we will confront the two languages using the philosophy of
the {\it quark-hadron duality} ---   that is how most of the  
results discussed in this section, a bunch of beautiful
{\it exact theorems} of QCD, will be obtained.

\subsection{QCD Inequalities. The Vafa-Witten Theorem}
As was discussed above, the octet of pseudoscalar mesons $(\pi, K, 
\eta)$ can be interpreted as 
that of the pseudo-Goldstone
particles appearing due to the  spontaneous 
chiral symmetry
breaking according to the pattern (\ref{patbreak}) in
the massless limit. That is the reason why 
the pseudoscalar mesons are lighter than
those with other quantum numbers.
It is interesting that the latter statement can be formulated as an exact 
theorem of QCD without any reference to the (experimental!) fact that the
chiral symmetry is broken.

Consider a QCD-like theory with at least 2 quark flavors and assume that 
these quarks (denote them by $u$ and $d$) have equal masses $m_u = m_d = m$.
Consider a set of Euclidean correlators
 \be
C_\Gamma(x,y) \ =\ \langle J^{\bar u d}(x)  J^{\bar d u}(y) \rangle_{\rm vac} 
\ ,
\label{setcorr}
 \ee 
where $ J^{\bar u d} (x)$ are flavor-changing bilinear quark currents
$ J^{\bar u d} = \bar u \Gamma d$ with Hermitean 
$$\Gamma = 1, \,\,\gamma^5,\,\,
i\gamma_\mu, \,\,\gamma_\mu \gamma^5,\,\, i\sigma_{\mu\nu}\,.$$
 At large distances, 
the correlators (\ref{setcorr}) decay exponentially 
 \be
\label{corrasmes}
C_\Gamma(x,y) \ \propto \ \exp\{-M_\Gamma|x-y| \}
  \ee
where $M_\Gamma$ is the mass of the lowest meson state in the corresponding 
channel.\footnote{We assume here that the quarks are confined, 
otherwise the whole
discussion is pointless.}

On the other hand, the correlators 
(\ref{setcorr}) of the quark currents can be
presented in the form
 \be
\label{TrGSGS}
C_\Gamma(x,y) \ =\ - Z^{-1} \int d\mu_A {\rm Tr} 
\left\{ \Gamma {\cal G}_A(x,y) \Gamma {\cal G}_A(y,x) \right\}\ ,
 \ee
where 
\be 
  \label{measQCD}
d\mu_A \ =\ \prod_{x\mu a} dA_\mu^a (x) \prod_f {\rm det}
\| m_f - i /\!\!\!\!{\cal D} \| 
\exp \left\{ - \frac 1{4g^2} \int F_{\mu\nu}^a   F_{\mu\nu}^a  d^4x \right \}
  \ee
is the standard QCD 
measure and ${\cal G}_A(x,y)$ is the Euclidean Green function
of the $u-$ and $d-$ quarks in a given gauge-field background. 
Note that, when writing down Eq.~(\ref{TrGSGS}), we used the fact that 
$J^{\bar u d}$ is not a singlet in flavor [otherwise, the disconnected
contribution $\propto {\rm Tr} \left\{ \Gamma {\cal G}_A(x,x) \right\}
 {\rm Tr} \left\{ \Gamma {\cal G}_A(y,y) \right\}$ would appear 
in the right-hand 
side].
In addition   the assumption $m_u = m_d$ 
was made [otherwise, we would have two 
different  Green's functions ${\cal G}^u_A(x,y) \neq {\cal G}^d_A(x,y)$].

An important nontrivial relation
 \be
\label{gSg}
\gamma^5 {\cal G}_A(x,y) \gamma^5 \ =\ {\cal G}_A^\dagger (y,x)
 \ee
holds. To understand it, write
the spectral decomposition for  ${\cal G}_A(x,y)$,
  \be
\label{speccont}
{\cal G}_A(x,y) \ =\ \langle \psi (x) \bar \psi (y) \rangle^A \ =\ \sum_{k}
\frac {u_k (x) u_k^\dagger (y) }{m - i \lambda_k}\ ,
 \ee 
(cf. Eq.~(\ref{speclat})). Using the symmetry 
 $u_k \to \gamma^5 u_k,\ \ \lambda_k \to - \lambda_k$, we obtain
 \be
 &&\gamma^5 {\cal G}_A(x,y) \gamma^5  =
\sum_{k}
\frac {[\gamma^5 u_k (x)][\gamma^5 u_k (y)]^\dagger }{m - i \lambda_k} 
\nonumber\\[0.1cm] 
&&=\ \sum_{p}
\frac {u_p (x) u_p^\dagger (y) }{m + i \lambda_p}=
\left[ \sum_{p}
\frac {u_p (y) u_p^\dagger (x) }{m - i \lambda_p}  \right]^\dagger =\, 
 {\cal G}_A^\dagger (y,x) \,,
 \ee
as annonced. We see that the pseudoscalar correlator
$$\sim \left \langle {\rm Tr} \{ \gamma^5 {\cal G}_A(x,y) 
\gamma^5 {\cal G}_A(y,x) \} \right
\rangle \ =\  \left \langle {\rm Tr} \{ | {\cal G}_A(x,y)|^2 \} \right
\rangle $$
plays a distinguished role -- it presents an absolute
upper bound for any other such correlator. The fastest way to show 
this is
to expand the $4 \times 4$ matrix ${\cal G}_A(x,y)$ over the full basis
 \be
\label{structS}
{\cal G}_A(x,y) &=& s(x,y) + \gamma^5 p(x,y) + i \gamma_\mu v_\mu(x,y) +
\gamma_\mu \gamma^5 a_\mu(x,y)  
\nonumber\\[0.1cm]
&+&\frac 12 i \sigma_{\mu\nu} t_{\mu\nu}(x,y)\ .
  \ee
Then
  \be
&&- \frac 14 C^A_{\gamma^5}(x,y) \ = \ 
\frac 14   {\rm Tr} \{ \gamma^5 {\cal G}_A(x,y) \gamma^5 {\cal G}_A(y,x) \} 
 \nonumber\\[0.1cm]
&&=\frac 14   {\rm Tr} \{ |{\cal G}_A(x,y)|^2 \} \ =\ 
|s|^2 + |p|^2 + |v_\mu|^2 + |a_\mu|^2 + 
 \frac 12 |t_{\mu\nu}|^2\ ,
   \ee
but, say
   \be
  && - \frac 14 C^A_{\bf 1}(x,y) \ \equiv \ 
\frac 14   {\rm Tr} \{  {\cal G}_A(x,y)  {\cal G}_A(y,x) \}
=\frac 14   {\rm Tr} \{ {\cal G}_A(x,y) \gamma^5 
{\cal G}_A^\dagger(x,y)  \gamma^5 \} \nonumber\\[0.2cm] 
&& =
|s|^2 + |p|^2 - |v_\mu|^2 - |a_\mu|^2 + \frac 12 |t_{\mu\nu}|^2\,  .
 \ee
The inequalities 
  \be
  \label{ineqPS}
|C^A_{\gamma^5}(x,y)| \geq |C_\Gamma^A(x,y)|
  \ee
 in any given gauge 
background, the
positivity of the measure (\ref{measQCD}) in Eq.~(\ref{TrGSGS}), 
and the asymptotics (\ref{corrasmes}) imply that the mass $M_{PS}$ of the 
(lightest) pseudoscalar meson in the $\bar u d$ channel is less or may be
equal to the masses $M_S,\, M_V,\, M_A,\, M_T$ of the lightest scalar, vector,
axial, and tensor states.\footnote{We 
were a little bit sloppy here. The  inequalities
(\ref{ineqPS}) hold strictly speaking only for 
tree correlators, not for renormalized ones. However, 
renormalization only brings about multiplicative    
factors, which do not depend on distance. 
Thus, taking $|x - y|$ to infinity {\it before} 
the limit $\Lambda_{\rm UV} 
\to \infty$ is done, we can ensure that the inequalities (\ref{ineqPS}) 
for renormalized correlators at large distances are  fulfilled.} 

Let us emphasize again that this statement can be justified only in 
the theory
with the positive measure (\ref{measQCD}) [e.g. in the theory 
with nonzero
vacuum angle $\theta \neq 0 $, the measure {\it is} not positive and 
pseudoscalar
states {\it need} not to be the lightest], with 
equal quark masses, and only for
the those states that are not flavor-singlet.
For flavor-singlet states it need not be true.  Consider, for example,
the theory with just one quark of a large mass. Then the lowest meson states
would be made of gluons and would know nothing about quarks. The 
lowest glueball state is believed to be scalar rather than  pseudoscalar.

In Sec.~1, we have 
mentioned already the  Vafa-Witten theorem saying that the
vector isotopic symmetry is not broken spontaneously in QCD. Now we are ready
to prove it. Indeed, if such a breaking occured, 
the massless Goldstone scalar 
particles would appear in the spectrum. The inequality $M_{PS} \leq M_S$
implies that a massless {\it pseudoscalar} particle would also exist. But
the theory with $m_u = m_d \neq 0$  (which duly enjoys the exact isovector
 symmetry,
a possible spontaneous breaking of which is under discussion now) 
has no exact axial isotopic symmetry, and there are {\it no reasons} 
for the massless pseudoscalar state to exist. So, it does not exist, hence 
the 
massless scalar does not exist either, and 
the isovector symmetry is not broken.

Many more inequalities of this kind (e.g.~ $M_N \geq M_\pi$ or $M_{\pi^+}
\geq M_{\pi^0}$) can be formulated, but their proof relies on
some extra assumptions.  We address the reader to
Ref. 10 for a nice recent review (see also S.
Nussinov's contribution  in this Volume).

\subsection{Euclidean Dirac Spectral Density}

Consider the Euclidean Dirac operator $/\!\!\!\!{\cal D}$ in a given gauge 
field background $A_\mu(x)$. We assume that the system is placed in a finite
4-volume so that the spectrum of $/\!\!\!\!{\cal D}$ is discrete. Let
$\{\lambda_k\}$ be the background-dependent set of eigenvalues of 
$/\!\!\!\!{\cal D}$. The {\it spectral density}\,\footnote{The 
notion of   spectral density and the definition (\ref{specdens})
are also widely used in  condensed matter physics and nuclear physics. It
is especially useful if a system is disordered or involves elements of 
disorder like it is the case for  electron spectra in most solids or for 
energy levels in complicated nuclei. It makes sense 
also for ordered systems (such as 
metals). In this case, rather than averaging
over stochastic external
field, one averages over some interval of eigenvalues
$\Delta \lambda$ much larger than characteristic level spacing, but much less
than a characteristic scale of $\lambda$ on
which $\rho(\lambda)$ is essentially
changed.} is defined as follows:
   \be
\label{specdens}
 \rho(\lambda) \ =\ \left\langle \frac 1V \sum_k \delta \left( \lambda -
\lambda_k[A_\mu(x)] \right)  \right\rangle \ ,
  \ee
where the average is done with the  weight 
function (\ref{measQCD}).
The $\gamma^5$ symmetry of the spectrum  implies that
$\rho(\lambda)$ is an even function of $\lambda$. 

In contrast to  solids or  nuclei, the spectral density (\ref{specdens}) 
is defined in the Euclidean space and seems to have no direct 
physical meaning. There are, however, a set of remarkable identities which
relate the spectral density of
the Euclidean Dirac operator to  physical 
observables. The simplest such identity relates    the spectral density
at ``zero virtuality'' $\lambda = 0$ to the quark condensate.

 To derive it, 
  set  $x=y$ in the spectral decomposition (\ref{speccont}), integrate it 
over $\frac 1V d^4x$,
and perform the averaging over the
gauge fields with  weight (\ref{measQCD}).
In view of the definitions (\ref{condfg}), (\ref{conddel}), 
and (\ref{specdens}),
using the symmetry $\rho(-\lambda) = \rho(\lambda)$, and assuming the reality 
of $\Sigma$, we obtain
 \be
 \label{Banks0}
\Sigma   \ =\ \left\langle \sum_k \frac 1{m - i\lambda_k} \right \rangle
\ =\ \int_{-\infty}^\infty \frac {\rho(\lambda) d\lambda }{m - i\lambda} 
\ =\ 2m \int_0^\infty  \frac {\rho(\lambda) d\lambda }{\lambda^2 + m^2} \ .
 \ee
To understand better this formula, let us look first what happens for free
fermions. As there is no physical dimensionfull scale in this case [remember
that $\lambda_k$ in Eq.~(\ref{specdens}) are eigenvalues of the {\it massless}
Dirac operator], $\rho(\lambda) = C\lambda^3$ on dimensional 
grounds. By 
counting the eigenvalues of the free Dirac operator
 \be
 \label{specfrefer}
 \lambda(n_\mu) \ =\ \frac{2\pi}L \sqrt{\sum_\mu \left(n_\mu + \frac 12 
\right)^2}
 \ee
[antiperiodic boundary conditions for the fermions in all 4 directions are 
chosen, $n_\mu$ are integer, and each level (\ref{specfrefer}) involves an 
extra $2N_c$-fold
 degeneracy] in the 4D ball $1/L\ll \lambda < \Lambda$, it is not
difficult to determine $C = N_c/(4\pi^2)$. Thus,
  \be
\label{densfree}
  \rho^{\rm free}(\lambda)\ =\ \frac {N_c}{4\pi^2} \lambda^3 \ .
  \ee
In  the interacting theory, the spectral density behaves as
$  \rho(\lambda)\ \propto \ \lambda^3$ for $\lambda$  much 
greater than the characteristic hadron scale $\mu_{\rm hadr}$ so that 
interaction is weak. To be more precise, the power $\lambda^3$ can be 
multiplied by an
anomalous dimension factor $$\sim \left(\ln \frac {\lambda }{\mu_{\rm hadr}} 
\right)^\alpha\,.$$ A recent one-loop calculation\,\cite{Zyab} 
implies  that $\alpha$ is nonzero and negative.

We see that the integral in Eq.~(\ref{Banks0})    
diverges quadratically in the
ultraviolet. The same result can be obtained directly by calculating the 
fermion bubble graph in the momentum representation
 \be
\label{condUV}
\langle \psi(0) \bar \psi (0) \rangle \ = \ \int \frac{d^4p_E}{(2\pi)^4}
 \ {\rm Tr}
\left\{ \frac{/\!\!\!{p_E} + m}
{p_E^2 + m^2} \right\} \ \propto \ m \Lambda_{UV}^2\ .
 \ee
Thus, strictly speaking, the formula (\ref{Banks0}) does not make much sense
as it stands. Note, however, that even though the (purely perturbative) 
contribution (\ref{condUV}) diverges in 
the ultraviolet, it vanishes in the 
chiral limit $m \to 0$. The whole point is that in QCD, the integral
(\ref{Banks0}) acquires an additional {\it nonperturbative} contribution
coming from the region of small $\lambda$ which survives in the 
``continuum chiral  thermodynamic limit'' ({\it first} $V \to \infty$, 
{\it then} $m \to 0$, and only then the ultraviolet cutoff is lifted
$\Lambda_{UV} \to \infty$). The fact that chiral symmetry is broken 
spontaneously {\it means} that the vacuum expectation value 
$ \langle \psi(0) \bar \psi (0) \rangle $ is nonzero in this 
particular limit.

Obviously, the necessary condition for the
condensate to develop is $\rho(0) \neq 0$. Neglecting all
 terms which vanish in the continuum  chiral  thermodynamic limit  defined
above, we obtain finally the famous {\it Banks-Casher 
relation}\,\cite{Banks}
 \be
\label{Banks}
 \langle \psi(0) \bar \psi (0) \rangle_{\rm vac}\ \equiv \ \Sigma \ =\ \pi 
\rho(0)\,.
  \ee
Note that the result does not depend on flavor which tells us again that the
flavor vector symmetry is not broken.

Not only $\rho(0)$, but also the form of $\rho(\lambda)$ at small $\lambda 
\ll \mu_{\rm hadr}$ can be determined.\cite{Stern} 
Consider the theory with $N_f \geq 2$
light quarks of common mass $m$. Let us study  an integrated 
correlator in this theory,
 \be
 \int d^4x \langle S^a(x) S^b(0) \rangle \ =\ \frac 1V \int d^4x d^4y
  \langle S^a(x) S^b(y) \rangle \ ,
  \ee
where $S^a(x) = \bar \psi(x) t^a \psi(x)$ and $t^a$ is the generator of the
SU$(N_f)$ flavor group. Fix a particular gluon background and define
  \be
\left.  C^{ab} \right|_A \ =\ - \frac 1V \int d^4x d^4y {\rm Tr}
\left\{ t^a  {\cal G}_A(x,y) t^b  {\cal G}_A(y,x) \right\}\ .
\label{CabA}
\ee
Substitute here the spectral decomposition (\ref{speccont}) for 
$ {\cal G}_A(x,y)$, do the integration and perform  averaging over 
the gluon
fields trading the sum over eigenvalues for the integral 
over the spectral
density (\ref{specdens}). We obtain
  \be
 \label{Cabint}
C^{ab} &=& - \frac{\delta^{ab}}{2V} \left\langle \sum_k 
\frac 1{(m - i\lambda_k)^2} \right \rangle\   = \  - \frac {\delta^{ab}}{2}
\int_{-\infty}^\infty \frac {\rho(\lambda) d\lambda }{(m - i\lambda)^2}
= \nonumber\\[0.2cm]  
&-&{\delta^{ab}} 
\int_{0}^\infty \frac {\rho(\lambda)(m^2 - \lambda^2)  }{(m^2 + \lambda^2)^2}
 d\lambda \ ,
 \ee
where the property $\rho(-\lambda) = \rho(\lambda)$ was used.

\begin{figure}
 \begin{center}
        \epsfxsize=200pt
        \epsfysize=150pt
        \vspace{-5mm}
        \parbox{\epsfxsize}{\epsffile{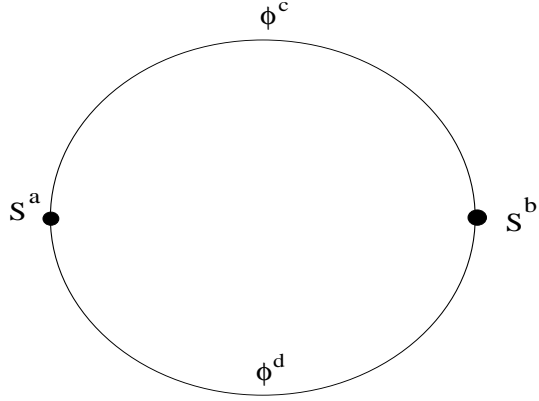}}
        \vspace{5mm}
    \end{center} 
\caption{Pseudogoldstone loop in  scalar correlator.}
\label{Sternfig}
\end{figure}

On the other hand, the same correlator can be saturated by physical states 
among which the pseudo-Goldstone states play a distinguished role. 
Consider the 1-loop graph in Fig.~\ref{Sternfig} describing the contribution
 of 
the 2-Goldstone intermediate state $\sim \int \langle 0|S^a |\phi^c \phi^d 
\rangle \langle \phi^c \phi^d | S^b | 0 \rangle $ (obviously, the one-particle
state does not contribute because the pseudo-Goldston
mesons are pseudoscalars
while $S^a(x)$ is scalar). To calculate it, we need to know the vertex
$\langle 0|S^a |\phi^c \phi^d \rangle $ which can be determined 
via the 
generating functional of QCD involving  scalar sources $u^a$ coupled to the
current $S^a$. 
Adding the source term $u^a S^a$ to the Lagrangian amounts 
to adding $u^a t^a$
to the quark mass matrix ${\cal M}$.  The latter enters also the mass term
(\ref{Lchirmass}) in the effective Lagrangian. Expanding $U$ up to the second
order in $\phi^a$ and varying it with respect to $u^a$, we obtain
 \be
 \label{Sfifi}
\langle 0|S^a |\phi^c \phi^d  \rangle \ =\  -\frac {\Sigma}{F_\pi^2} 
d^{abc}\,.
 \ee
The vertex is nonzero only for three or more flavors. 

Now we can calculate the graph in Fig.~\ref{Sternfig}. Actually, 
we cannot because
the integral diverges logarithmically in the
ultraviolet, but anyway  the
effective theory is not valid at high momenta (technically, the divergence is 
absorbed into local counterterms of higher order in $p^{\rm char} $ and $m$).
Only the infrared-sensitive part of the integral is relevant. A simple 
calculation gives
 \be
 \label{Cabinfr}
\left( C^{ab} \right)^{\rm infrared} \ =\ - \frac{N_f^2 - 4}{32 \pi^2 N_f}
\left( \frac \Sigma {F_\pi^2} \right )^2  \delta^{ab} \ln \frac
{M_\phi^2}{\mu_{\rm hadr}^2} \ .
  \ee
Compare it now with Eq.~(\ref{Cabint}). Note first of all that the constant 
part $\rho(0)$ does not contribute here,
$$ \int_0^\infty \frac {m^2 - \lambda^2}{(m^2 + \lambda^2)^2} d\lambda \ =\ 0 
\ . $$
Thus, only the difference $\rho(\lambda) - \rho(0)$ is relevant. It is easy
to see that, in order to reproduce the singularity $\sim \ln M_\phi^2 \sim
\ln m$, we should have  $\rho(\lambda) - \rho(0) = C|\lambda|$ at small
 $|\lambda|$. Substituting it in Eq.~(\ref{Cabint}) 
and comparing the coefficient of $\ln m$ with 
the coefficient of $\ln M_\phi^2$ in Eq.(\ref{Cabinfr}), we
finally obtain$\,$\cite{Stern}
 \be
\label{SmStern}
 \rho(\lambda) \ =\ \frac \Sigma \pi + \ \frac{N_f^2 - 4}{32 \pi^2 N_f}
\left( \frac \Sigma {F_\pi^2} \right )^2 |\lambda| \ + o(\lambda)\,.
 \ee
Thus, for $N_f \geq 3$, the spectral density has a nonanalytic ``dip'' at
$\lambda = 0$. The behavior is smooth in the theory with two light flavors.
Physically, it is rather natural that the more is number of flavors, the 
stronger is the suppression of 
$\rho(0)$. The determinant factor in the measure (\ref{measQCD})
punishes small eigenvalues, and the larger is $N_f$, the more important is
this factor. By ``analytic continuation'' of this argument, one should expect
a nonanalytis bump rather than a nonanalytic dip at $\lambda = 0$ in the 
case $N_f = 1$. Indeed, Eq.~(\ref{SmStern}) displays such a bump. 
One should not forget, of course, that the whole derivation was based on the 
effective chiral Lagrangian approach
and does not directly apply to the case $N_f = 1$.
Some additional, more elaborate reasoning shows, however, 
that a bump at $N_f = 1$ as predicted by Eq.~(\ref{SmStern}) 
is there.\cite{Toublan} 
It is also confirmed by a numerical calculation in the instanton liquid 
model.\cite{Verb}
 
\subsection{Chiral Symmetry Breaking and Confinement}

Look again at Eq.~(\ref{chiranom}). The axial current entering
the left-hand side 
is an external current in a sense that no dynamical field is
coupled directly to $j_{\mu5}$. But the fields entering on the right-hand
 side are
dynamical  gluon fields present in the QCD Lagrangian. 

In the chiral (left-right asymmetric) gauge theories like the standard 
electroweak model, both vector and axial currents are coupled directly to
the physical gauge fields. Anomalous divergence of 
such current would mean explicit
breaking of the gauge invariance which is not nice. Therefore, in chiral
theories, one should 
always take care that such purely {\it internal anomalies} would 
cancel out at
the end of the day. 
In the Standard Model, they do.

Let us discuss, however, purely {\it external anomalies} in QCD which are not
related to  breaking of any symmetry but just mean that certain correlators
involving external currents are singular.\footnote{The conventional 
anomaly (\ref{chiranom}) is
related to  the correlator (\ref{T3point}) involving 
both internal ($j_\mu^a$) and external ($j_\lambda^5$) currents and
can be called {\it mixed} in this setting.}

As a simplest nontrivial example, consider the theory with two
 massless flavors 
and look at the correlator
  \be
 \label{KabH}
K_{\mu\nu}^{AB,\ {\cal H}} (q) \ = \ i \int \langle T\{j_{\mu 5}^A(x) 
j_\nu^B(0) \} \rangle_{\cal H} \ e^{iqx} d^4x\ ,
 \ee
where $A,B$ are flavor indices and ${\cal H}$ is the external 
flavor-singlet ``magnetic field.''\,\footnote{The 
quotation marks distinguish ${\cal H}$ which couples to the 
baryon charge from the physical magnetic
field which has the matrix structure diag$(2/3, -1/3)$ and is a mixture of 
isotriplet and isosinglet. But we are not interested in dynamics of 
electromagnetic
or weak currents here. In QCD proper, all color-singlet currents are external.
 ${\cal H}$
is just a source of such vector flavo-singlet current.}
The correlator (\ref{KabH}) is nothing but a three-point vacuum expectation
value (\ref{T3point}) in  kinematics in which one of 
the external momenta 
associated with the vector current is set to zero.

The one-loop calculation of the corresponding graph displays  a
 singularity,
 \be
\label{Kabres}
 K_{\mu\nu}^{AB,\ {\cal H}} (q) \ = \ 
 - \frac {{\cal  H}}{2\pi^2} \frac {q_\mu 
\tilde{\epsilon}_{\nu\alpha} q^\alpha}{q^2} \cdot N_c \cdot \frac 12 
\delta^{AB}\ .
 \ee
The imaginary part of this amplitude is also singular,
$\sim \delta(p^2)$, which can be related to the masslessness of 
quarks.\cite{Dolgov}
However, the quarks (in contrast to electrons in QED) do not exist as physical
particles  due to confinement and one can ask where does the singularity
 in
the imaginary part Im $ K_{\mu\nu}^{AB,\ {\cal H}} (q) \propto \delta(q^2)$ 
come from?
This is a good question, the answer is  better still: the singularity
$\sim 1/q^2$ comes from the propagator of a massless Goldstone boson, which
appears due to the spontaneous 
chiral symmetry breaking  and which is directly
coupled to the axial current $j_{\mu 5}$ (see the middle graph in 
Fig.~\ref{Hooftfig}).

\begin{figure}
  \begin{center}
        \epsfxsize=300pt
        \epsfysize=100pt
        \vspace{-5mm}
        \parbox{\epsfxsize}{\epsffile{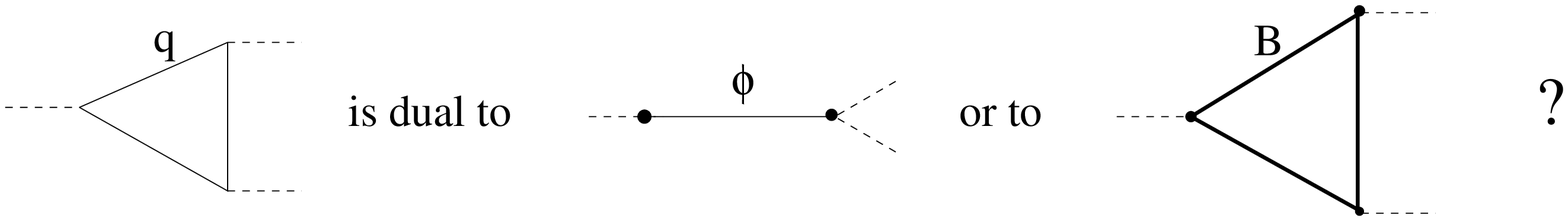}}
        \vspace{5mm}
    \end{center} 
\caption{Saturating the external anomaly.}
\label{Hooftfig}
\end{figure}

Let us ask now:  can one reproduce the singularity in  
Eq.~(\ref{Kabres}) {\it without } Goldstone bosons and without spontaneous
chiral symmetry breaking, but in some other way ?

As far as the theory with two
 light quarks is concerned, the answer is positive:
the singularity of the correlator above can,  be reproduced,
in principle, if 
{\it massless baryons} are present. Proton and neutron represent, like quarks,
a flavor SU$(2)$ doublet. There are $N_c = 3$ quark doublets and only one
baryon doublet $|P\rangle = |uud\rangle$ and   $|N\rangle = |udd\rangle$. 
The absence of the overall
$N_c$ factor is compensated, however, by the fact that the { baryon charge}
of nucleon is 3 times larger than that of  the quark, and the vertex involving
the ``magnetic field `` ${\cal H}$ is 3 times larger for 
 baryons.

Thus,  this purely algebraic   {\it anomaly matching} argument due to
 't Hooft\,\cite{Hooft} does not rule out a dynamical
scenario where the physical spectrum in the theory with just two massless quark
flavors would not involve massless pions, but, instead, the 
massless proton and neutron. 
It is rather remarkable that, in the theories with $N_f \geq 3$, the scenario
with massless baryons {\it is } ruled out. Suppose that, instead of the octet
of massless Goldstone fields, we have an octet of massless baryons. The 
contribution of the corresponding triangle graph in (\ref{KabH}) would have
the same structure as in Eq.~(\ref{Kabres}), but with the factor
 \be
{\rm Tr} \ \{T^a T^b \} \ =\ C_8 \delta^{ab}
 \ee
instead of $\delta^{ab}/2$, where $T^a$ are the flavor generators in the octet
representation. To find the {\it Dynkin index} of the octet representation
$C_8$, it 
is sufficient to assume that $a=b=1$ and decompose the octet with respect
to the SU$(2)$ flavor subgroup: ${\bf 8} = {\bf 3} + {\bf 2} + {\bf 2} + 
{\bf 1}$. The contribution of each doublet to $C_8$ is  $1/2$ and the 
contribution of the 
triplet is $2$. Adding it together, we obtain $C_8 = 3 \neq 
1/2$
and the required result (\ref{Kabres}) {\it is} not reproduced. Also 
a massless decuplet and all other possible color-singlet baryon 
representations would give the coefficient 
in front of  the singularity much larger than
that in Eq.~(\ref{KabH}), and the anomaly matching condition would not be 
fulfilled. Therefore, massless baryons do not exist.\footnote{ To be 
quite precise, one could, in  principle,  saturate
the anomaly with {\it several} baryon multiplets with positive and negative
baryon charges. This possibility is so unaesthetic, however, that it can be 
rejected by that reason.}

We have arrived at a remarkable 
result. In QCD with three massless quarks, the 
assumption of confinement allowing the existence of only colorless states in
the physical spectrum {\it and} the anomaly matching condition lead to the
conclusion that massless Goldstone states {\it must} appear and chiral symmetry
{\it should} be spontaneously broken. If the latter is not true, the only
possibility to saturate the anomaly is to assume that massless quarks still
exist as physical states in the spectrum and there is no confinement!

In the real world, confinement and spontaneous 
chiral symmetry breaking 
in the limit of massless $u$, $d$,
and $s$ quarks are experimental facts. 
Whether or not these 
phenomena 
take place  in hypothetic theories with $N_f \geq 4$  is an open
question. It is quite possible that starting, say, from $N_f \stackrel{?}= 
6 $, the small
eigenvalues in the Dirac operator spectrum are punished 
 by the 
determinant   factor so strongly that $\rho(0) =0$ and, in view of the Banks-Casher 
relation (\ref{Banks}), the quark condensate vanishes, and the symmetry is not
 broken. The   anomaly matching argument tells then that there is no 
confinement in this case.\footnote{We know, 
of course, that if the number of the quark flavors is {\it
very} high $N_f > 16$, the asymptotic freedom is lost and we cannot expect
confinement. We are almost sure that quarks and gluons are not confined
at $N_f = 16$ or $N_f = 15$, in which case the theory is asymptotically free, 
but has an infrared
fixed point at a small value
of $\alpha_s$,
and the coupling constant never grows large.\cite{Zaks}
There are some reasons to believe, however,
 that confinement is lost at a  smaller value of $N_f$, not just $N_f = 15$.}  

We have called this result --
that the chiral symmetry breaking and confinement go 
together --  remarkable.
It is also somewhat misterious. Even though we do not
understand well dynamical reasons for confinement to occur, we still expect
that {\it the same} mechanism which works for the theory with $N_f=3$ works
also for the theory with $N_f = 2$. But  't Hooft's argument works only
for $N_f \geq 3$...  
 
\section{Mesoscopic QCD}

To make the path integral finite-dimensional and the spectrum of the Dirac
operator discrete, not only an ultraviolet lattice
regularization should be introduced, but one also has to consider the theory
in a finite volume. If the volume is very large, much larger that the pion
 Compton wavelength, the dynamics of the theory is basically the same as
that in the continuum limit. However, in practical calculations
characteristic lattice sizes are not too large, and finite volume effects
are essential. It is therefore important to estimate theoretically these 
effects.
In  this section, we derive a set 
of {\it exact} relations for
QCD placed in a finite Euclidean box.\cite{LS}
  They rely heavily on the presence
of the chiral symmetry and on the notion of quark-hadron duality
used in the previous section.
 We will always assume that the size of the box $L$
is large $L \gg \mu_{\rm hadr}^{-1}$, but  the requirement
$L \gg M_\pi^{-1}$ is not imposed.

\subsection{Partition Function: $N_f = 1$}

Consider first the theory with just one light quark flavor. 
We assume that the quark mass $m$ 
is complex 
so that the mass term in the Lagrangian has the form
  \be
 \label{massNf1}
{\cal L}_m^{N_f = 1} \ =\ m \bar \psi_R \psi_L\  +\  \bar m 
\bar \psi_L \psi_R\ ,
 \ee
and the 
vacuum angle $\theta$  is nonzero. The partition function of the theory
can be written in the form
  \be
 \label{Zexten}
Z(\theta)\ =\ \exp\{ - V \epsilon(\theta, m, L) \}\ ,
  \ee
 where $V = L^4$ and $\epsilon(\theta, m, L)$ 
tends to the vacuum energy
density $\epsilon_{\rm vac}(\theta, m)$ in the 
thermodynamic limit $L \to \infty$. 
It is important that the finite volume corrections
$\epsilon(\theta, m, L) - \epsilon_{\rm vac}(\theta, m)$ are exponentially 
suppressed $\propto \exp \{- \mu_{\rm hadr} L\}$. Indeed, the corrections 
appear due to modification of the spectrum of excitations in the finite
box by the Casimir mechanism. At the technical level, they are due to   
modification of the corresponding Green  functions by finite volume effects.
But in the theory with $N_f = 1$, all physical (meson and baryon) excitations
are massive and their Green functions decay as
$ \exp \{- \mu_{\rm hadr} R\}$ at large distances. Also the modification of the
Green  functions (at coinciding points) due to boundary effects has the order
$ \exp \{- \mu_{\rm hadr} L\}$.

The vacuum energy depends on 
the quark mass $m$ and vacuum angle $\theta$. It turns
out that $\epsilon_{\rm vac}$ is actually a function of just one complex
parameter which is $m e^{i\theta}$. To see that, express $Z(\theta)$ as a 
series
 \be
 \label{ZtetQCD}
 Z(\theta)  & \sim &  \sum_{q = -\infty}^\infty e^{-iq \theta}
 \int \prod_{\tau, \vec{x}} dA_\mu^a(\tau, \vec{x}) 
\det \left\| -i /\!\!\!\!{\cal D}  + m \frac {1 - \gamma^5}2 + \bar m
\frac {1 + \gamma^5}2  \ \right\| \nonumber \\[0.2cm]
&\times & \exp \left\{
 - \frac 1{4g^2} \int_{0}^{\beta}  d\tau \int d\vec{x}
 \left[ (F_{\mu\nu}^a)^2 \right] \right \} \ \equiv 
\ \sum_{q = - \infty}^\infty e^{-iq \theta} Z_q  \ ,
   \ee 
where $Z_q$ is given by  the functional integral over 
the quark and gluon fields in
the sector of definite topological charge $q$. Consider some particular
$Z_q$ and  suppose
for definiteness that $q > 0$.
One of the factors entering the 
 integrand
of $Z_q$ is the quark determinant,
  \be
 \label{detqprod}
&&\det \left\| -i /\!\!\!\!{\cal D}  + m \frac {1 - \gamma^5}2 + \bar m
\frac {1 + \gamma^5}2  \ \right\| \, =  \bar m^q  \prod'_k  \det \left \|
\begin{array}{cc} \bar m & -i \lambda_k \\ -i \lambda_k & m  \end{array} \right\| 
 \nonumber \\[0.2cm]
&&=\bar m^q \prod'_k (\lambda_k^2 + \bar m m )\ ,
  \ee
where  the product
$\prod'$ runs over  nonzero $\lambda_k$'s, and the factor $\bar m^q$
reflects the presence of $q$ right-handed fermion zero modes as the index
theorem (\ref{AtSing}) requires [cf. Eq.~(\ref{det0n0}) written earlier for the 
case of real masses]. 
 If $q < 0$, the factor $\bar m^q$ should  
be substituted by $m^{-q}$ . Averaging (\ref{detqprod}) over the gluon field
configurations with a given $q$  and substituting it in Eq.~(\ref{ZtetQCD}),
we obtain that each term in the sum depends, indeed, only on the combinations
$z = m e^{i\theta}$ and $\bar z$ (being real, the partition function and
$\epsilon_{\rm vac} \propto \ln Z$ cannot, of course, be holomorphic in $z$).

Next we assume that  $|m| \ll \mu_{\rm hadr}$, that the dependence of 
$\epsilon_{\rm vac} $ on mass is analytic,
and expand
  \be
 \label{Enexptet}
\epsilon_{\rm vac} \ =\ \epsilon_0 - \Sigma {\rm Re} \left( m e^{i\theta} 
\right ) \ + o(m) \ ,
\ee
  where $\Sigma$ is a real constant which, for $\theta = 0$ and real $m$,
coincides with the variation of the vacuum energy with respect to mass, i.e. with 
the quark condensate\,\footnote{A finite 
nonzero value of the quark condensate was exactly
our assumption when we   wrote  the expansion (\ref{Enexptet}).} 
$ \langle \psi \bar\psi \rangle_{\rm vac}$. 
Substituting (\ref{Enexptet}) in Eq.~(\ref{Zexten}) and
comparing it with the Fourier expansion (\ref{ZtetQCD}), we can derive
 \be 
 \label{ZqFour}
Z_q \ \propto \ \frac 1{2\pi} \int_0^{2\pi} d\theta e^{iq\theta}
e^{V \Sigma m \cos \theta + o(m)} \ =\ I_{q} (m\Sigma V)
  \ee
[$I_q(x) = I_{-q}(x)$ stands for the Bessel functions 
which grow  exponentially  at
large real $x$]. 

It is very instructive now to consider two limits: {\it (i)} 
$m\Sigma V \gg 1$ and {\it (ii)} $m\Sigma V \la 1$ (keeping the assumption
$L\mu_{\rm hadr} \gg 1$). The limit {\it (i)} is a genuine thermodynamic
limit. The sum (\ref{ZtetQCD}) is saturated by many terms with large values of
$|q|$. Actually, one easily derives (for any $m\Sigma V$) 
 \be
 \label{hiNf1}
 \langle q^2 \rangle_{\rm vac} \ =\ m \Sigma V\,.
  \ee
The quantity $\chi = \langle q^2 \rangle_{\rm vac} /V$ can also be 
 expressed via the
correlator of topological charge densities,
 \be
 \label{chidef}
\chi \ =\  \int   \ d^4x \ \left\langle  \frac 1{16\pi^2}  
 {\rm Tr} \{F_{\mu\nu} \tilde F_{\mu\nu}\} (x) \ \ 
\frac 1{16\pi^2}  
 {\rm Tr} \{F_{\mu\nu} \tilde F_{\mu\nu}\} (0) \right\rangle \,,
 \ee
and is called the  topological
susceptibility of the vacuum. Taking a double variation of
Eq.(\ref{ZtetQCD}) with respect to $\theta$, we 
 relate it to the curvature of the function $\epsilon_{\rm vac} (\theta)$
 at $\theta = 0$:
  \be
\label{susceptpend}
   \chi \,  =\left. - \lim_{V \to \infty} \frac 1{V Z(0)}
 \frac {\partial^2 Z(\theta)}
  {\partial \theta^2}  \right|_{\theta = 0} \, =\, \left.
 \frac {\partial^2 \epsilon_{\rm vac} (\theta)}
  {\partial \theta^2}  \right|_{\theta = 0}  \  .
  \ee
The region 
\be
\label{mesosc}
\mu_{\rm hadr}^{-1} \ \ll\ L \ \la \ (m\Sigma)^{-1/4}
  \ee
can be called {\it mesoscopic}. The word is borrowed from the condensed matter
physics where mesoscopic system is defined as a system involving a large
enough number of particles for the statistical description to be adequate, but
which is not sufficiently large for the boundary effects to be irrelevant. The
 same
is true in our case. In particular, when $m\Sigma V \ll 1$, the partition 
function is well approximated by just the first term $q = 0$ and, if we are
interested in the mass dependence, --- by the terms $q = \pm 1$ corresponding
to the instanton and anti-instanton sectors.

If  $m\Sigma V \ll 1$, the fermionic condensate is provided by instanton-like
configurations supporting a single fermion zero mode. If  $m\Sigma V \gg 1$,
the condensate appears due to finite average density of small but nonzero
eigenvalues of the Dirac operator, according to the Banks-Casher formula
(\ref{Banks}). Quite remarkably, the {\it value} of the condensate is 
the same, however, and does not depend on the parameter $m\Sigma V$.

Note that, if $m$ is real and $\theta = 0$ (more generally, if the
combination
$\theta_{\rm phys} = \theta + {\rm arg}(m)$ on which everything depends 
vanishes), all   terms in the partition function (\ref{ZtetQCD}) are 
positive
and the path integral for $Z(\theta)$ has a probabilistic interpretation.
And if not --  then not.

\subsection{Partition Function: $N_f \geq 2$}

The spectrum of the theory with several light flavors involves light
pseudo-Goldstone states. If the length of our box is comparable with the 
Compton wavelength of pseudo--Goldstone particles, finite volume corrections
in $\epsilon(\theta, m, L)$  in Eq.~(\ref{Zexten}) are not exponentially small
and should be taken into account. We will be interested here in the region
 \be
 \label{mesNf}
\mu_{\rm hadr}^{-1} \ \ll\ L \ \ll \frac 1{M_\pi} \sim \frac
1{\sqrt{m\mu_{\rm hadr}}} \ .
 \ee
The scale $\sim m^{-1/4} \mu_{\rm hadr}^{-3/4}$ entering Eq.~(\ref{mesosc})
and lying in the middle of the newly defined mesoscopic interval (\ref{mesNf})
is relevant   in the multiflavor case too. The formulae we are going to derive
in this section will be universally valid, however, for any value of 
$m\Sigma V$ as long as the condition (\ref{mesNf}) is satisfied. 

  The main idea is to present the partition function as the path integral
in the {\it effective theory} (\ref{Leffchir2}), (\ref{Lchirmass}) and notice
that, when the condition $LM_\pi \ll 1$ is fulfilled, only the zero Fourier
harmonics of the pseudo--Goldstone fields are relevant.\footnote{One can 
observe it by
looking at the pseudo-Goldstone Green   function in finite volume,
$$
G_\pi(x) \ =\ \frac 1V \sum_{\{n_\mu\}} \frac {e^{ipx}}{M_\pi^2 + \left(
\frac {2\pi n_\mu}L \right)^2} \,.  $$} 
 Thus, we can disregard the kinetic term, and the path integral is reduced 
to an ordinary one. We want to calculate the partition function at arbitrary
$\theta$ and need to know the form of the effective potential at arbitrary
$\theta$. 
In the case $N_f = 1$ all physical
quantities depended on the combination $me^{i\theta}$, but not on $m$ and 
$\theta$ separately.
By the same token, for $N_f \geq 2$, they depend on the combination
${\cal M} e^{i\theta/N_f}$ (where ${\cal M}$ is a complex mass matrix).
The proof is quite analogous and is based on the multiflavor version of
Eq.~(\ref{detqprod}),
  \be
 \label{detqprodN}
&&\det \left\| -i /\!\!\!\!{\cal D}  + {\cal M} \frac {1 - \gamma^5}2 + \ 
{\cal M}^\dagger
\frac {1 + \gamma^5}2  \ \right\|\ =\  \left[ \det\|{\cal M}^\dagger\|  \right]^q  
 \nonumber\\[0.2cm] 
 & & \times \prod'_k  \det \left \|
\begin{array}{cc} {\cal M}^\dagger & -i \lambda_k \\ -i \lambda_k & 
{\cal M}  \end{array}  \right\|  =\, 
\left[ \det\|{\cal M}^\dagger\| 
\right]^q   \prod'_k \det \|\lambda_k^2 + {\cal M}^\dagger {\cal M} \|\, .
  \ee
Similar to the case $N_f = 1$, we see that the physical vacuum angle on which 
all physical quantities depend is 
 \be
\label{tetfizNf}
\theta_{\rm phys} \ =\ \theta + {\rm arg} (\det \| {\cal M} \| )\,,
 \ee
rather than just $\theta$. In particular, the statement made earlier
 that the vacuum angle in QCD is either zero or very small refers to the
combination (\ref{tetfizNf}).

  Thus, we can write
\be
\label{partNtet}
Z_{N_f} (\theta) \ =\ \int dU \exp \left\{ V \Sigma {\rm Re}
\left[ {\rm Tr} \{ {\cal M} 
 e^{i\theta/N_f} U^\dagger \} \right]
\right\}\ ,
  \ee
where the integral is done over the group SU$(N_f)$ with the proper Haar
measure. Integrating it further over $d\theta$ as in Eq.~(\ref{ZqFour}), we
find the partition function in the sector with a definite topological
charge $Z_q$. In the simple case ${\cal M} = m {\bf 1}$, a beautiful analytic
expression for $Z_q$ can be derived,
 \be
 \label{LSdet}
Z_q \ =\ \det \left \|
\begin{array}{cccc} 
I_{q} (\kappa) & I_{q+1} (\kappa) & \ldots & I_{q + N_f - 1} (\kappa) \\
 I_{q-1} (\kappa)&  I_{q} (\kappa) & \ldots &  I_{q+N_f - 2} (\kappa) \\
\vdots & \vdots & \vdots & \vdots \\
 I_{q - N_f + 1} (\kappa)& \ldots & \ldots &  I_{q} (\kappa)  \end{array}
\right \|\ ,
  \ee
where $\kappa = |m\Sigma V|$.

Let us assume $\theta = 0$ , $m$ real, and let us study the quantity
 \be
 \label{condmV}
 \langle \psi \bar \psi \rangle _{m, V} \ =\ \frac 1{N_f V} \frac {\partial }{\partial m}
\ln Z \ ,
 \ee
 the quark condensate at finite mass and finite volume. In the 
limit $\kappa \to \infty$, the
 integral in Eq.~(\ref{partNtet}) has the main support in the region 
$U \sim 1$, and we
 obtain\,\footnote{Note 
the presence of the power-like preexponential factor in Eq.~(\ref{ZNas}).  It reflects 
the fact
that,  while the condition  $LM_\pi \ll 1$ is satisfied and higher 
Fourier harmonics of the pion
fields are decoupled, we are not in the true thermodynamic limit yet. 
In the latter, the
partition function is given by a simple exponential as in Eq.~(\ref{Zexten}),
 and the finite-volume
corrections to $\epsilon(\theta, m, L)$ are suppressed as
$\exp\{-M_\pi L \}$. }
  \be
  \label{ZNas}
Z_{N_f} \ \propto \ \frac {e^{N_f \kappa}}{\kappa^{(N_f^2 - 1)/2} }\,,
  \ee
  so that the condensate (\ref{condmV}) tends to a constant $\Sigma$ as it should.
  
In the opposite limit $\kappa \ll 1$, $Z_{q \neq 0}$ are suppressed compared to $Z_0$
 as $\kappa^{|q|N_f}$ reflecting the presence of $N_f|q|$ fermion zero modes in the sector 
with the topological charge $q$. The condensate  (\ref{condmV}) is also 
suppressed $\propto \kappa$. This is due to the fact that, in the multiflavor case, the
condensate plays the role of the order parameter signaling the
 spontaneous chiral symmetry breaking.
Strictly speaking, spontaneous symmetry breaking never occurs in quantum systems at
finite volume  in the 
 sense that the true ground state wave function is always symmetric  even 
though the gap
separating it from  excited asymmetric states may be very (exponentially)
small.\footnote{Never say never: the statement above is not correct for supersymmetric 
systems.\
  But QCD is not supersymmetric and we may forget this subtlety.} 
So , for $N_f > 1$, the condensate should vanish in the chiral limit $m \to 0$ no matter
  how large the fixed volume $V$ is. And it does.  

\subsection{Spectral Sum Rules}

Let us return to a   simpler case, $N_f = 1$. Assume that $m$ is real and $q$
is positive, and rewrite the result (\ref{ZqFour}) as 
\be 
\label{predvq}
\left \langle  m^q \prod'_k (\lambda_k^2 + m^2) \right \rangle_q
\ =\ C  I_{q} (\kappa) \ ,
  \ee
where the averaging is done over the gluon field configurations with a given
topological charge $q$. We can expand now the two sides of Eq.~(\ref{predvq})
in $m$ and compare the coefficients of the expansion. The first nontrivial
relation  is obtained when
comparing the terms $\propto m^q$ and $\propto m^{q+2}$ . We obtain
  \be
 \label{LSq2}
\left\langle \!\! \left\langle \sum'_k \frac 1{\lambda_k^2} \right 
\rangle \!\! 
 \right \rangle_q  \ =\ \frac {(\Sigma V)^2}{4(q+1)} \ ,
 \ee
where the sum runs over the nonzero eigenvalues and the 
symbol $\langle \! \langle  \cdots \rangle \! \rangle$ means 
averaging 
with  weight $\exp\{ - S_g \} \prod'_k \lambda_k^2$ (the second factor
is the fermionic determinant  in the limit $m
\to 0$, with the discarded  zero modes).

Expanding Eq.~(\ref{predvq}) up to the order $\sim m^{q+4}$, we obtain the
following sum rule: 
  \be
 \label{LSq4}
\left\langle \!\! \left\langle \sum'_{k \neq l} \frac 1{\lambda_k^2 \lambda_l^2} 
\right \rangle \!\! 
 \right \rangle_q  \ =\ \frac {(\Sigma V)^4}{16(q+1)(q+2)} \ .
 \ee
Similar relations can be derived in the multiflavor case. 
Expanding Eq.~(\ref{LSdet}) in mass
and comparing the terms 
$\sim m^{N_f q},\ \sim m^{N_f q + 2}$ , and $\sim m^{N_f q+4}$, we obtain the 
same relations (\ref{LSq2}), (\ref{LSq4}),
with $q$ being substituted by $q + N_f - 1$. We can  extract more
information by considering the expression for the partition function with {\it 
different} quark masses. Two different sum rules can be derived by considering
  the terms $\sim m^{N_fq+4}$ of the expansion,
   \be
 \label{LSq4N}
\left\langle \!\! \left\langle \left(\sum'_{k} \frac 1{\lambda_k^2 } \right)^2
\right \rangle \!\! 
 \right \rangle_q  \ &=&\ \frac {(\Sigma V)^4}{16\left[(q+N_f)^2 - 1\right] } \ ,
\nonumber \\[0.2cm]
\left\langle \!\! \left\langle \sum'_{k} \frac 1{\lambda_k^4 } 
\right \rangle \!\! 
 \right \rangle_q  \ &=&\ \frac {(\Sigma V)^4}{16(q+ N_f)\left[(q+N_f)^2 - 1 
\right]} \ .
 \ee
 Subtracting one sum rule
in Eq.~(\ref{LSq4N}) from  the other and setting $N_f = 1$, we reproduce 
Eq.~(\ref{LSq4}), but the relations (\ref{LSq4N}) are valid only for $N_f \geq 2$. 
Actually, 
the average $\langle \! \langle \sum_k' (1/\lambda_k^4) 
\rangle \! \rangle_q$ diverges at $N_f = 1$. For $N_f = 2$, there are two 
sum rules at each level.  For $N_f = 3$, three different sum rules can be 
derived 
starting from  the terms $\sim m^{N_fq+6}$ of the expansion, etc.

The sum rules (\ref{LSq2}) can be presented as the integrals
 $$
V \int \ \frac {\rho(\lambda)}{\lambda^2} \ d\lambda \ ,$$
where $\rho(\lambda)$ is the microscopic spectral density (\ref{specdens}) 
in the sector with a given topological charge $q$..
The sum rules  (\ref{LSq4}) and the first sum rule in Eq.~(\ref{LSq4N})
are not expressed in terms of  $\rho(\lambda)$ only, but also via a certain 
integral
involving the correlation function $\rho(\lambda_1, \lambda_2)$. The sum rule
for $\langle \! \langle \sum_{k \neq l \neq p}' (1/\lambda_k^2 \lambda_l^2 
\lambda_p^2)  \rangle \! \rangle_q$ is expressed in  
 terms of an integral of the 
correlator  $\rho(\lambda_1, \lambda_2, \lambda_3)$ , etc. 
Using ingenious  arguments which are beyond the scope of our discussion
here, J. Verbaarschot and collaborators  managed to determine
the functional form of $\rho(\lambda)$ and of all relevant correlators.
For example,\cite{Zahed} 
  \begin{eqnarray}
\label{VZT}
\rho_{N_f, q}(\lambda) &=& \frac {\Sigma^2 V |\lambda|}{2}
\left[ J^2_{N_f + |q|} (\Sigma V \lambda) \right.\nonumber\\
&-&\left.  
 J_{N_f + |q| + 1} (\Sigma V \lambda)  J_{N_f + |q| - 1} (\Sigma V \lambda)
\right]\,.
  \end{eqnarray}
The expression is valid for $ \lambda \ll \mu_{\rm hadr}$, and the function
is ``alive'' in the region of very small eigenvalues
$\lambda \sim 1/\Sigma V$. In the thermodynamic limit $\lambda \Sigma V
\to \infty$, the function (\ref{VZT}) tends to the constant $\Sigma/\pi$
in agreement with the theorem (\ref{Banks}).

Spectral sum rules are well adapted to be confronted with the numerical 
lattice calculations. The main interest here is not so much to ``confirm'' 
 these
exact theoretical results  by computer, but, rather, to test lattice methods.
This was a challenge for lattice people for some time. By now reasonably
 good numerical data were obtained,   they agree well with the theory.\cite{latLS}
 One
can expect that the accuracy will grow substantially if 
the algorithms based on the lattice
fermion action with exact chiral   symmetry 
(as was  discussed   in detail  in   Sec.~2)
are implemented.

\section{QCD at finite $\theta$}

We know that in the real world the gauge group is $SU(3)$ and 
we have 6 fundamental quarks with some particular mass values.
 We also know that
$\theta_{\rm phys} $ defined in (\ref{tetfizNf}) 
is very close to zero. It presents a considerable 
theoretical interest, however, to study what happens in the same theory, 
but with other values of the parameters. Though these ``fairy''\,\footnote{This 
word was
coined by chess problemists. A  fairy chess problem is a 
problem referring to a game with
modified rules or on a nonstandard board.}
 variants
of the theory do not have a direct relation to reality, in order to
understand {\it well} the physics of our world, it is important to understand
also how it varies if the rules of the game are changed. 

One of the ways to modify the theory is to assume a nonzero 
$\theta_{\rm phys}$. We will not  assume that the quark masses coincide
with their experimental values, but will keep them (or rather some of them)
small enough for the chirality considerations to be relevant. 
The expressions for the partition function in  finite
volume  as a function of $\theta$ were written in Sec.~3. In this section, we
will concentrate, however, not on the mesoscopic regime $m\Sigma V \sim 1$, 
but on  dynamics of the theory in the thermodynamic limit   
 $m\Sigma V \gg 1$.

The vacuum energy at finite $\theta$ (its mass-dependent part) is 
obtained$\,$\cite{dVW} 
by minimizing the effective chiral potential over $U$,
  \be
  \label{Evactet}
\epsilon_{\rm vac}(\theta) \ =\ -\Sigma\ \min_U \left[ {\rm Re \ Tr }\left\{
{\cal M} e^{i\theta/N_f} U^\dagger \right\} \right]\ .
 \ee
In the two-flavor case, a nice analytic expression for
$\epsilon_{\rm vac}(\theta)$
can be written for a generic mass matrix ${\cal M}$,
   \be
   \label{EvacN2}
   \epsilon_{\rm vac}(\theta) \ =\ -\Sigma \sqrt{{\rm Tr} \{ {\cal M}^\dagger
 {\cal M}
   \} + e^{i\theta} \det \|{\cal M}\| + e^{-i\theta} \det \|{\cal M}^\dagger\|
 }\ .
    \ee
In the general case  it is a continuous smooth function of $\theta$. It is, 
of course, also periodic: $\epsilon_{\rm vac}(\theta + 2\pi)  = 
\epsilon_{\rm vac}(\theta)$. The case of degenerate masses is more subtle. 
Substituting ${\cal M} = m {\bf 1}$ (with
 real $m$) in Eq.~(\ref{EvacN2}) we obtain
     \be
     \label{Evac }
\epsilon_{\rm vac}(\theta) \ =\ -2m\Sigma \left| \cos \frac \theta 2  \right| \
 .
 \ee
 This function is still periodic in $\theta$, but is obviously singular at 
$\theta = \pi$. 
 A mathematical reason for this is rather simple. The function (\ref{EvacN2}) 
has two
 stationary points: a  minimum and a  maximum. At $\theta = \pi$, they fuse 
together.
 The maximum and minimum of energy are plotted in Fig.~\ref{minimaxfig}.
 We see that
 two smooth curves $\propto \pm \cos (\theta/2) $ cross  at the point 
$\theta = \pi$ and,
 after passing this point, the maximum and minimum are interchanged. While
  $\epsilon_{\rm vac}(\theta)$ is determined by the curve $-2m\Sigma 
\cos (\theta/2)$ at $ \theta < \pi$, it is determined by the
 curve $2m\Sigma  \cos (\theta/2)$ for $\theta > \pi$.  
  
 \begin{figure}
 \begin{center}
        \epsfxsize=300pt
        \epsfysize=200pt
        \vspace{-5mm}
        \parbox{\epsfxsize}{\epsffile{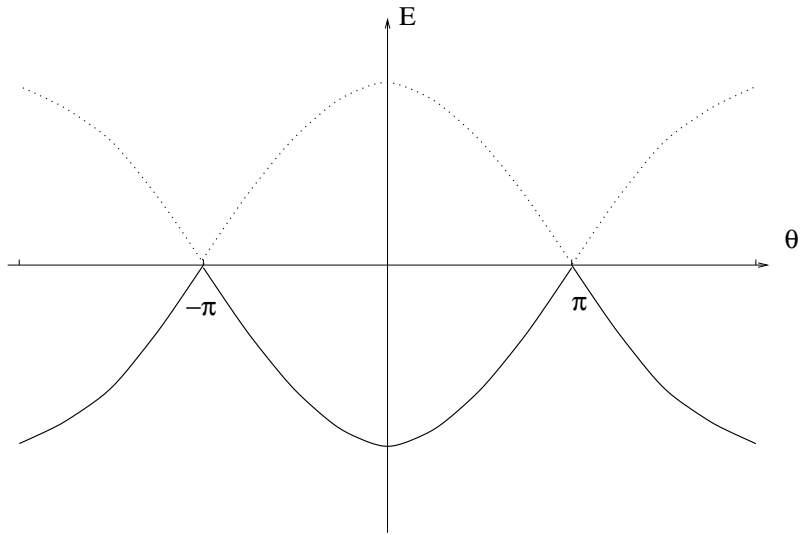}}
        \vspace{5mm}
    \end{center}
\caption{$N_f = 2$: Minima (solid line) and maxima (dotted line) of the effective potential
for different $\theta$.}
\label{minimaxfig}
  \end{figure}

 Physically, this is the situation of the second order phase transition: 
the energy is a continuous
  function of the parameter $\theta$, but its derivative is not. A more 
detailed analysis
  shows, however, that the conclusion of the existence of the second order 
phase transition
  is an artifact of the approximation   used. Indeed, for $N_f = 2$, the 
potential in 
  Eq.~(\ref{Evactet}) just vanishes identically at the point $\theta = \pi$. 
That means that
 the pion excitations become exactly massless and chiral 
symmetry is restored.\footnote{Another way to 
understand it is to look at the Gell-Mann-Oakes-Renner
  relation (\ref{GMOR}) . Equation (\ref{GMOR}) was
originally written   for 
$\theta = 0$, but a
  generalization for arbitrary $\theta$ is rather straightforward. 
In our case, we even do
  not need it: the point $\theta = \pi,\ m_u = m_d$ is equivalent to the point
  $\theta = 0,\ m_u = - m_d$ by virtue of (\ref{tetfizNf}), and we see that 
the pion mass vanishes.}  
     
It is clear from physical considerations that, no matter what the value
  of $\theta$ is, the symmetry (\ref{SULSUR}) is broken explicitly if the 
quarks are massive.  For $\theta = \pi$  the leading term vanishes, 
but the breaking is still there  being induced by 
the terms of {\it higher order in mass} in the chiral effective potential.\cite{tetpi} 
The relevant term has the structure $\propto m^2\sin^2 (\theta/2){\rm Tr} \{U^2\}$ . 
  
One can show that taking this into account modifies the curves in 
Fig.~\ref{minimaxfig} in the vicinity of $\theta = \pi$ so that, 
at the point $\theta = \pi$,  one  has two exactly
  degenerate vacuum states separated by a low but nonvanishing barrier.  If 
$\theta$ is  slightly less than $\pi$, the degeneracy is lifted, and 
we have a metastable vacuum state on top of the true vacuum. When $\theta$ slightly
exceeds $\pi$, the  roles of these
  two states is reversed: the one which was stable before becomes metastable,  
and the other way around. And this is exactly the physical picture of the 
{\it first order} phase transition with superheated water, supercooled 
vapor and all other familiar  paraphernalia. 
  
 Consider now the theory with  three degenerate light flavors.
By a conjugation
$U \to RUR^\dagger$, any unitary matrix $U$ can be rendered diagonal, 
 $$U = {\rm diag}
(e^{i\alpha}, e^{i\beta}, e^{-i(\alpha + \beta)} )\,.$$ When ${\cal M} = m
{\bf 1}$, the conjugation does not change the potential 
in Eq.~(\ref{Evactet}). For diagonal $U$, the latter acquires the form
 \be
 \label{Uab}
V(\alpha, \beta) \ =\ -m\Sigma \left[ \cos \left( \alpha - \frac \theta 3
\right) + \cos \left( \beta - \frac \theta 3
\right) + \cos \left( \alpha + \beta + \frac \theta 3
\right) \right]\,.
  \ee
 The function $U$ has six  stationary points,
\be
\label{statp}
&&{\rm \bf I} :\ \alpha = \beta = 0, \ \ 
{\rm \bf II} :\ \alpha = \beta = -\frac {2\pi}3, \ \ 
{\rm \bf III} :\ \alpha = \beta =  \frac {2\pi}3 \nonumber \\[0.2cm]
&&{\rm \bf IV}: \  \alpha = \beta = -\frac {2\theta}3 + \pi,\ \ 
{\rm \bf IV}a: \  \alpha = -\alpha - \beta =  -\frac {2\theta}3 + \pi, 
\nonumber \\[0.2cm]
&&{\rm \bf IV}b: \  \beta = -\alpha - \beta = -\frac {2\theta}3 + \pi\,.
  \ee
The points {\bf IV}a and {\bf IV}b are obtained from  {\bf IV} by
the  Weyl
permutations $\alpha \leftrightarrow \beta, \ 
\alpha \leftrightarrow -\alpha - \beta$, etc, 
and their physical properties are the same. Actually, we have
here not 3 distinct stationary points, but, rather,  a   four-dimensional manifold
SU$(3)/[{\rm SU}(2) \otimes {\rm U}(1)]$ of the physically equivalent stationary
points related to each other by  conjugation. The values of the 
potential at the stationary points are
  \be
  \label{En}
&&\epsilon_{\rm \bf I} = -3m\Sigma \cos \frac \theta 3,\ \ 
\epsilon_{\rm \bf II} = -3m\Sigma \cos  \frac {\theta + 2\pi} 3, \nonumber \\[0.2cm]  
&&\epsilon_{\rm \bf III} = -3m\Sigma \cos \frac {\theta - 2\pi} 3,\ \ 
 \epsilon_{\rm \bf IV} = m\Sigma \cos  \theta \,.
  \ee
Studying  the expressions (\ref{En}), and the matrix of  the second 
derivatives of the potential (\ref{Uab}) at $\theta = \pi$, one can readily 
see that {\it i)} the points 
{\bf I} and
{\bf III} are   degenerate minima separated by a barrier; 
{\it ii)} the point {\bf II} is a 
maximum, and {\it iii)} the points {\bf IV} are  saddle points.
The physical picture we arrive at is that of the first order phase transition. 
The situation is even simpler than in the two-flavor case since 
the barrier between the minima appears   in the leading order
and   the  subleading $ m^2$  terms  in the potential is of no concern 
to us in the case at
hand.

  \begin{figure}
 \begin{center}
        \epsfxsize=300pt
        \epsfysize=200pt
        \vspace{-5mm}
        \parbox{\epsfxsize}{\epsffile{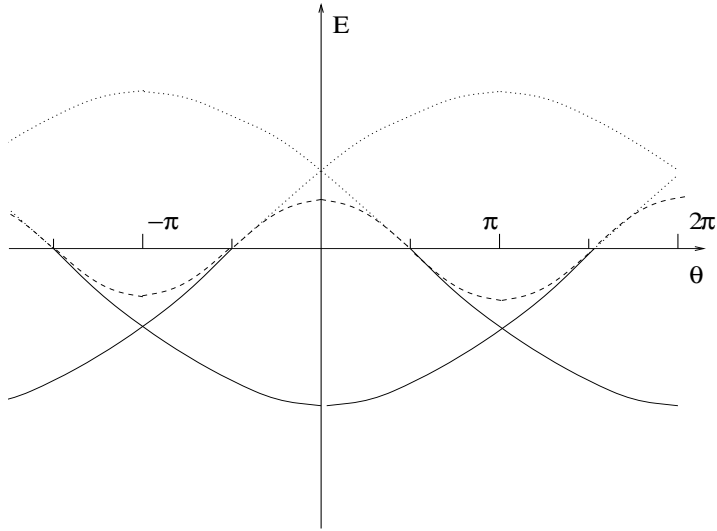}}
        \vspace{5mm}
    \end{center}
\caption{$N_f = 3$: Stationary points of $V(\alpha, \beta)$ for different 
$\theta$. 
The solid lines are  minima,  the dotted lines are  maxima, and the dashed 
line are  saddle points}
\label{statfig}
  \end{figure}

Figure \ref{statfig} illustrates how the stationary points of the potential   
move  when the vacuum angle is changed. At $\theta = \pi/2$,
$\theta = 3\pi/2$, etc, the metastable  minima coalesce with the saddle points and
 disappear. No trace of them is left at $\theta = 0$.
One can show that also for the physical
values of the quark masses the metastable
vacua are absent at $\theta = 0$.
 
 If the quark masses are not degenerate but their values are close, 
nice curves in Fig.~\ref{statfig}
 are distorted a little bit, but  physics remains largely the same: first 
order phase transition
 is robust and cannot be destroyed by small variations of parameters. 
If the masses are
 essentially different, the phase transition may disappear. 
This is true, in particular, for physical
 values of the quark masses. In this case, the vacuum energy depends on
 $\theta$ analytically, as in Eq.~(\ref{EvacN2}). Unfortunately, no such simple formula for 
 $\epsilon_{\rm vac}(\theta)$ can be written for a generic 
mass matrix ${\cal M}$ for three or more
 flavors.
 
 What we can do easily and quite generally is to determine the behavior of the function 
  $\epsilon_{\rm vac}(\theta)$ at small $\theta$ and the topological 
susceptibility (\ref{susceptpend}).
  Let us do it for arbitrary $N_f$ still assuming, for simplicity, that ${\cal M}$ is real and
diagonal.
 Then  $U$ can  be taken to be  diagonal too,
  $$ U \ =\ {\rm diag} \left(e^{i\alpha_1}, \ \ldots \  e^{i\alpha_{N_f}}\right),\ \ \ \ \ \ 
  \sum_{j = 1}^{N_f} \alpha_j \ =\ 0 \ .$$
  For $\theta \ll 1$, we anticipate that the vacuum values of $\alpha_j$ will also be small.
  We can write the effective potential as
   \be
   \label{VeffNf}
   V^{\rm eff}(U) &=&   - \Sigma \sum_{j = 1}^{N_f}  
m_j \cos \left(\alpha_j - \frac \theta {N_f} \right) \nonumber \\[0.2cm]
&=& {\rm const} + \frac \Sigma 2  \sum_{j = 1}^{N_f}  m_j\left( \alpha_j 
 - \frac \theta {N_f}  \right)^2 + \cdots 
  \ee
  Adding here a Lagrange multiplier 
term $\lambda \sum_{j = 1}^{N_f} \alpha_j $ and
minimizing
  the expression thus obtained  over $\alpha_j$ and $\lambda$, we find
   \be
   \label{alphaj}
   \alpha_j \ =\ \theta \left[ \frac 1{N_f} - 
\frac 1{m_j    \sum_{j = 1}^{N_f}  m_j^{-1}}  \right] \ .
    \ee
    Substituting this result for $\alpha_j$
 in  Eq.~(\ref{VeffNf}), we finally obtain\,\cite{dVW}
     \be
     \label{hiQCDNf}
 \chi = m_{\rm eff}^{-1} = \ \frac {\partial^2 {\cal E}_{\rm vac}(\theta)}
  {\partial \theta^2} \left. \right|_{\theta = 0} = \frac 1V \langle q^2 
\rangle_{N_f} \, =\, 
  \Sigma \left[ \frac 1 {m_1} + \cdots +   \frac 1 {m_{N_f}} \right]^{-1} \, .
    \ee  
    Roughly speaking, $\chi$ is proportional to the lighest quark mass 
(lighest quark masses).
    In actual QCD 
     \be
     \label{hiQCD}
     \chi_{\rm QCD} \ =\ \frac{m_u m_d}{m_u + m_d} \Sigma\ ;
     \ee
 the strange and other quarks are too heavy to be relevant.
 If ${\cal M} = m {\bf 1}$,
     Eq.~(\ref{hiQCDNf}) is reduced\,\cite{Crewther} 
to $\chi = m\Sigma/N_f$. 
The latter result can be  easily reproduced by substituting the 
asymptotics (\ref{ZNas}) for $Z_q$ in the definition
$$ \langle q^2 \rangle \ =\ \frac {\sum_q q^2 Z_q}{\sum_q Z_q} \ .$$

\section*{References}

\end{document}